\documentclass[prd,twocolumn,showpacs,nofootinbib]{revtex4}

\usepackage{aas_macros}
\usepackage{amsfonts}
\usepackage{amsmath}
\usepackage{amssymb}
\usepackage{subfig}
\usepackage{graphicx}
\usepackage{url}
\usepackage{color}

\newcommand{\incgraph}[3]{\includegraphics[angle=#1, width=#2\textwidth]{#3}}

\begin{document}

\title{Accelerated Bayesian model-selection and parameter-estimation in continuous gravitational-wave searches with pulsar-timing arrays}

\author{Stephen Taylor}
\email[email: ]{stephen.r.taylor@jpl.nasa.gov}
\affiliation{Jet Propulsion Laboratory, California Institute of Technology, Pasadena, California 91109, USA}
\affiliation{Institute of Astronomy, University of Cambridge, Madingley Road, Cambridge, CB3 0HA, UK}
\author{Justin Ellis}\thanks{Einstein fellow}
\affiliation{Jet Propulsion Laboratory, California Institute of Technology, Pasadena CA 91109, USA}
\affiliation{Center for Gravitation and Cosmology, Department of Physics, \\ University of Wisconsin-Milwaukee, P.O. Box 413, Milwaukee, Wisconsin 53201, USA}
\author{Jonathan Gair}
\affiliation{Institute of Astronomy, University of Cambridge, Madingley Rd., Cambridge, CB3 0HA, UK}

\date{\today}


\begin{abstract}
We describe several new techniques which accelerate Bayesian searches for continuous gravitational-wave emission from supermassive black-hole binaries using pulsar timing arrays. These techniques mitigate the problematic increase of search-dimensionality with the size of the pulsar array which arises from having to include an extra parameter per pulsar as the array is expanded. This extra parameter corresponds to searching over the phase of the gravitational-wave as it propagates past each pulsar so that we can coherently include the pulsar-term in our search strategies. Our techniques make the analysis tractable with powerful evidence-evaluation packages like \textsc{MultiNest}. We find good agreement of our techniques with the parameter-estimation and Bayes factor evaluation performed with full signal templates, and conclude that these techniques make excellent first-cut tools for detection and characterisation of continuous gravitational-wave signals with pulsar timing arrays. Crucially, at low to moderate signal-to-noise ratios the factor by which the analysis is sped up can be $\gtrsim 100$, permitting rigorous programs of systematic injection and recovery of signals to establish robust detection criteria within a Bayesian formalism.


\end{abstract}


\maketitle


\section{Introduction}

The last several years have seen a growing effort to develop robust and powerful data-analysis techniques for the purpose of detection and characterisation of gravitational-waves (GWs) using ensembles of precisely timed Galactic millisecond pulsars. When a GW passes the Earth-pulsar line of sight, it causes a perturbation to the intervening space-time metric which may leave a measurable imprint on the time-of-arrival (TOA) of radio pulses from regularly observed millisecond pulsars \citep{burke-1975,sazhin-1978,detweiler-1979,estabrook-1975}. With a pulsar timing array (PTA) \citep{foster-backer-1990} we effectively create a Galactic-scale GW detector, sensitive in the $\sim 1-100$ nHz band. 
There are three separate PTA efforts underway: the European Pulsar Timing Array (EPTA) \citep{eptareview2013}, the Parkes Pulsar Timing Array (PPTA) \citep{parkesreview2013} and the North American Nanohertz Observatory for Gravitational waves (NANOGrav) \citep{nanogravreview2013}. There are also ongoing efforts to combine the techniques and data from all three PTAs within the umbrella consortium of the International Pulsar Timing Array (IPTA) \citep{iptareview2013}.

The current focus of PTA searches is to uncover evidence for a nanohertz stochastic GW background, most likely composed of many inspiraling supermassive black-hole (SMBH) binary signals overlapping in the frequency-domain which cannot be resolved separately \citep{rajaromani1995,jaffe-backer-2003,wyithe-loeb-2003}. While this background may dominate at the lowest detectable frequencies (where the characteristic strain is expected to be largest) at higher frequencies the stochasticity of the signal begins to break down, and in individual Monte Carlo realisations of SMBH binary populations we see single bright sources rising above the level of the unresolved background to become the dominant signal \citep{sesana-vecchio-colacino-2008,sesana-vecchio-volonteri-2009,ravi-2012}. It stands to reason then that several massive nearby binaries may be bright enough to resolve with PTAs, presenting a unique opportunity to probe the very early inspiral regime of their coalescence, and thereby offering a complementary probe of the massive black-hole population to eLISA/NGO \citep[e.g.,][]{sesana-review-2013-1,wenjenet2011}. 

The earliest attempts to constrain the properties of single resolvable sources with PTAs focused on nearby candidate systems. \citet{lommenbacker2001} investigated the level of timing-residuals expected from a binary system in Sgr A$^*$, finding that such a system would be beyond the sensitivity of near-future observations, while other nearby systems may offer a better chance of hosting a detectable binary. A much lauded result of pulsar-timing analysis was when the nearby radio galaxy 3C 66B was ruled-out as hosting a $1.05$ year orbital-period\footnote{Alarm bells always ring in pulsar-timing analysis when periodicities close to $1$ year appear, since a necessary step involves converting topocentric TOAs to barycentric TOAs.} SMBH binary system at greater than $95\%$ confidence \citep{jenet2004_3C66B}.

Techniques to infer the presence of the expected periodic TOA-deviations induced by a binary source have included both frequentist and Bayesian approaches. Due to the irregular sampling of pulsar TOAs, methods which have implemented power spectral summing \citep{yardley-2010} or ``harmonic summing'' \citep{jenet2004_3C66B} have used a Lomb-Scargle periodogram to avoid undesirable spectral leakage. We can also maximise our likelihood statistic over nuisance amplitude parameters to form the $\mathcal{F}$-statistic \citep{Fstat-paper-1998}, which has been applied to the detection of nearly-periodic signals in LIGO/Virgo/GEO data \citep[e.g.,][]{Fstat_virgo2014,Fstat2004,Fstat2007}, in the eLISA band \citep[e.g.,][]{Fstat_LISA2007}, and more recently in the nanohertz-sensitive PTA band \citep{babak-sesana-2012,ellisoptimal2012}. Time-domain techniques are now the favoured approach, and it has been realised that coherently including the ``{\it pulsar-term}'' contribution to the timing-residuals from when the GW passed the pulsar is hugely important for detection, sky-localisation, and distance determination \citep{corbin-cornish-2010,lee-wex-2011}. 

This pulsar-term arises when we integrate the response of pulsar-timing measurements to a GW over the path of the photons, giving contributions to the TOA deviations from either end of the Earth-pulsar timing baseline. 
The {\it Earth-term} adds coherently, but in previous analyses the pulsar-terms have been ignored as a form of self-noise whose contributions sum incoherently from separate pulsars. However, coherently including the pulsar-term can be regarded as the temporal equivalent of aperture synthesis \citep{corbin-cornish-2010}, increasing the baseline of PTA observations by thousands of years, and hence allowing us to track the orbital evolution of binary sources via the imprint of the GW in each distinct pulsar. Full Bayesian parameter estimation and evidence techniques now exist which include the pulsar-term by searching over each pulsar distance \citep{ellisbayesian2013}. However these typically require significant computational resources to explore the large-dimensional parameter space, and highly-tuned search algorithms to ensure phase coherence when searching over the distance. We side-step these issues by presenting fast techniques designed for a rapid first-analysis of the data, returning Bayes factor and parameter-estimation results which are in good agreement with full searches.

This paper is arranged as follows. In Sec.\ \ref{sec:signalmodel} we review the theory of timing-residuals induced by single resolvable GWs, along with templates to search for binaries which may or may not be evolving over the Earth-pulsar light travel-time. We also introduce our techniques, based on marginalising over the phase variables from each distinct pulsar, thereby collapsing the dimensionality of searches and accelerating evidence recovery. In Sec.\ \ref{sec:singlesourceresults} we compare the results of our model-selection with full searches, and investigate any potential biases in our parameter estimation. We state our conclusions in Section \ref{sec:singlesourceconclusions}.

In the following we define $G=c=1$.


\section{The signal} \label{sec:signalmodel}






The transverse-traceless (TT) gauge GW-tensor can be described as a linear superposition of ``plus'' and ``cross'' polarisation modes, with associated polarisation-amplitudes, $h_{\{+,\times\}}$, and basis-tensors, $e_{ab}^{\{+,\times\}}(\hat\Omega)$. In the context of single-source searches, the direction of GW-propagation, $\hat\Omega$, is written as $\left[-(\sin\theta\cos\phi)\hat{x} - (\sin\theta\sin\phi)\hat{y} - (\cos\theta)\hat{z}\right]$ such that $(\theta,\phi) = (\pi/2 - {\rm DEC}, {\rm RA})$ denotes the sky-location of the source in spherical polar coordinates.

As the GW propagates between the Earth and pulsar it creates a perturbation in the metric which causes a change in the proper distance to the pulsar, which in turn leads to a shift in the perceived pulsar rotational frequency. This fractional frequency shift of a signal from a pulsar in the direction of unit vector $\hat{p}$, induced by the passage of a single GW propagating in the direction of $\hat\Omega$ is \citep{anholm-2009,brook-flanagan-2011},
\begin{equation}
z(t,\Omega) = \frac{1}{2}\frac{\hat{p}^a\hat{p}^b}{1+\hat\Omega\cdot\hat{p}}\Delta h_{ab}(t,\Omega),
\end{equation}
where $\Delta h_{ab}\equiv h_{ab}(t_e,\hat\Omega) - h_{ab}(t_p,\hat\Omega)$ is the difference in the metric perturbation evaluated at time $t_e$ when the GW passed the solar system barycentre (SSB) and time $t_p$ when the GW passed the pulsar.
From simple geometrical arguments, we can write $t_p = t_e - L(1+\hat\Omega\cdot\hat{p})$, where $L$ is the distance to the pulsar. The integral of this {\it redshift} over time gives the GW contribution to the recorded pulse TOA. 
Consequently, this means that the timing-models which have been constructed to describe deterministic contributions to the pulsar TOAs (e.g.,\ quadratic spindown) will be slightly mismatched because we have not factored in the influence of GWs. This effect is observed in the {\it timing-residuals} which are the difference between the raw measured TOAs and the best-fit deterministic timing-model. These residuals encode the influence of noise and all unmodelled phenomena which influence the pulsar TOAs. The pulsar timing-residuals induced by a {\it single} GW source can be written as,
\begin{equation} \label{eq:GWinducedresiduals}
s(t,\hat\Omega) = F^+(\hat\Omega)\Delta s_+(t) + F^\times(\hat\Omega)\Delta s_\times(t)
\end{equation}
where $\Delta s_A(t) = s_A(t_p) - s_A(t_e)$, and $t_{\{p,e\}}$ denote the times at which the GW passes the pulsar and the Earth, respectively. The functions $F^A(\hat\Omega)$ are ``{\it antenna pattern}'' functions encoding the geometrical sensitivity of a particular pulsar to a propagating GW, defined as,
\begin{equation}
F^A(\hat\Omega)\equiv \frac{1}{2}\frac{\hat{p}^a\hat{p}^b}{1+\hat\Omega\cdot\hat{p}}e^A_{ab}(\hat\Omega).
\end{equation}

SMBH binaries are the primary candidate for nanohertz GWs. The population in this band are typically massive ($\gtrsim 10^8 M_{\odot}$), and in the early, adiabatic inspiral portion of their orbital evolution. Assuming circular orbits, the typical orbital velocity of these systems scales as \citep{corbin-cornish-2010},
\begin{equation}
v \simeq 2.5\times 10^{-2}\left(\frac{f}{10^{-8}\;{\rm Hz}}\right)^{1/3}\left(\frac{M}{10^8 M_{\odot}}\right)^{1/3},
\end{equation}
such that we are dealing with only mildly-relativistic binaries, with $v<<1$. Hence, the influence of BH-spin on the GW signal, which modifies the waveform at $1.5$ pN ($\propto v^3$), will be completely negligible for PTA observations, while orbital plane precession due to spin-orbit coupling may only be a consideration for the Square Kilometre Array (SKA) \citep{sesana-vecchio-measuring-2010,mingarelli2012}. Preliminary assessments of the importance of binary eccentricity indicate that the majority of the GW power will remain confined to the harmonic at twice the binary orbital frequency \citep{sesana-vecchio-measuring-2010}, however there is a growing concern that couplings between a binary and its environment can induce significant eccentricity, which may require this parameter to be included in waveform templates \citep{sesana-review-2013-2,tong2013,ravi_eccentricity_2014}. We ignore this effect here, and concentrate on circular, non-spinning SMBH binaries.

The periodically varying pulsar timing-residuals induced by a SMBH binary are derived from the quadrupolar waveform, and have the form \citep{wahlquist-1987,corbin-cornish-2010,ellisoptimal2012},
\begin{align} \label{eq:splusscross}
s_+(t) &= \frac{\mathcal{M}^{5/3}}{D_L\omega(t)^{1/3}}\left[-\sin\left[2\left(\Phi(t)-\phi_n\right)\right]\left(1+\cos^2\iota\right)\cos2\psi\right. \nonumber\\
&\left.- 2\cos\left[2\left(\Phi(t)-\phi_n\right)\right]\cos\iota\sin2\psi\right] \nonumber\\
s_\times(t) &= \frac{\mathcal{M}^{5/3}}{D_L\omega(t)^{1/3}}\left[-\sin\left[2\left(\Phi(t)-\phi_n\right)\right]\left(1+\cos^2\iota\right)\sin2\psi\right. \nonumber\\
&\left. + 2\cos\left[2\left(\Phi(t)-\phi_n\right)\right]\cos\iota\cos2\psi\right],
\end{align}
where $\psi$ is the GW polarisation angle; $\iota$ is the binary orbital-inclination angle; $\phi_n$ is the orbital phase at the line of nodes (defined as the intersection of the orbital plane with the tangent plane of the sky \citep{wahlquist-1987}); $\mathcal{M}=(m_1m_2)^{3/5}/(m_1+m_2)^{1/5}$ is the binary chirp mass defined in terms of the individual SMBH masses $m_{\{1,2\}}$; and $D_L$ is the luminosity distance to the source. Note that the chirp mass and orbital frequency are defined in terms of their observed values, where the rest-frame values are given by $\mathcal{M}_r=\mathcal{M}/(1+z)$, $\omega_r = (1+z)\omega$, and $z$ is the cosmological redshift. In a Universe with zero curvature the luminosity distance is defined in terms of the radial comoving distance $D_c$ by $D_L=(1+z)D_c$.

The rate of change of the binary orbital frequency due to GW emission is,
\begin{equation}
\dot{\omega} = \frac{96}{5}\mathcal{M}^{5/3}\omega^{11/3},
\end{equation}
with which we can derive the orbital frequency and phase,\footnote{We assume the phase evolution is driven entirely by GW emission.}  at a given time, $t$,
\begin{align} \label{eq:FullFreqPhase}
\omega(t) &= \omega_0\left(1 - \frac{256}{5}\mathcal{M}^{5/3}\omega_0^{8/3}(t-t_0)\right)^{-3/8},\\
\Phi(t) &\equiv \int_{t_0}^t\omega(t')dt' = \Phi_0 + \frac{1}{32\mathcal{M}^{5/3}}\left(\omega_0^{-5/3} - \omega(t)^{-5/3}\right).
\end{align}

The characteristic chirp timescale of an inspiraling binary is \citep{ellisbayesian2013},
\begin{equation}
\tau_{\rm chirp} \sim \frac{\omega_0}{\dot\omega_0} = 3.2\times 10^5\;{\rm yr}\left(\frac{\mathcal{M}}{10^8M_{\odot}}\right)^{-5/3}\left(\frac{f_0}{10^{-8}\;{\rm Hz}}\right)^{-8/3},
\end{equation}
which shows us that frequency and amplitude chirping of the binary over the course of typical PTA observation times ($10-20$ years) will be negligible compared to PTA frequency resolution ($\sim 1/T$) \citep{yardley-2010,sesana-vecchio-measuring-2010}, and can be safely ignored. Hence, we are looking for essentially monochromatic signals, and as such the {\it Earth-term} orbital frequency and phase are,
\begin{equation}
\omega_e(t) = \omega_0,\quad \Phi_e(t) = \Phi_0 + \omega_0(t-t_0).
\end{equation}

The corresponding variables for the {\it pulsar-term} must take into account the fact that the GW imprints a snapshot of the binary's orbital evolution as it passes {\it each} pulsar. As such, we deal with the {\it retarded time} $t_p$ which causes the pulsar-term to differ in phase from the Earth-term (and all other pulsar-terms) even if there is negligible frequency evolution over the Earth-pulsar light travel-time (highly unlikely). Frequency chirping is a long timescale effect for these systems. Indeed the value of $(\omega_0 - \omega(t_p))/\omega_0$ for a $10^8M_{\odot}$ chirp mass binary with $\omega_0=10^{-7}$ Hz and $L(1+\hat\Omega\cdot\hat{p})=2$ kpc is $\sim 0.03$. For the highest-mass system considered in this work ($7\times 10^8 M_{\odot}$ chirp mass binary with $\omega_0=2\pi\times 10^{-8}$ Hz, and most pulsars satisfying $L(1 + \Omega\cdot\hat{p})\leq 1$ kpc) the fractional difference between the Earth- and pulsar-term frequencies is $< 10\%$.
Hence we can Taylor-expand Eq.\ (\ref{eq:FullFreqPhase}) (and ignore evolution over the PTA observation window) to give,
\begin{align}
\omega_p(t) &\simeq \omega_0 - \dot\omega_0L(1+\hat\Omega\cdot\hat{p}),\nonumber\\
\Phi_p(t) &\simeq \Phi_{p,0} + \omega_0(t-t_0) - \dot\omega_0L(1+\hat\Omega\cdot\hat{p})(t-t_0),
\end{align}
where $L$ is the pulsar distance. The constant term $\Phi_{p,0}$ denotes the initial binary orbital phase of the pulsar-term, and is defined as
\begin{equation}
\Phi_{p,0} = \Phi_0 + \frac{1}{32\mathcal{M}^{5/3}}\left(\omega_0^{-5/3} - \omega_p^{-5/3}\right).
\end{equation}
We note that $\omega_p(t)$ is always less than or equal to the Earth-term frequency, such that a coherent measurement of the pulsar-term would afford an insight into the history of the binary's evolution.

We now have all the definitions we need to construct signal templates describing the pulsar-timing residuals induced by either a non-evolving or evolving SMBH binary. In all of the following we collect $\Phi_0$ and $\phi_n$ into one constant {\it initial phase} variable, $\phi_0 = \phi_n - \Phi_0$.

\subsection{Non-evolving template} \label{sec:NonEvolveTemplate}

Consider the low-frequency (or low chirp-mass) regime, where evolution of the source frequency is small, such that the frequencies of the GW when it passes the pulsar and the Earth are approximately the same. We can include the pulsar-term in our single-source template by modelling the signal in a single pulsar as the sum of two sinusoids of different phases. The signal template in the $\alpha^{\text{th}}$ pulsar is \citep{ellisoptimal2012},
\begin{equation}\label{eq:non-evolve-template}
s_{\alpha} = \sum_{i=1}^2 a_{i\alpha}(\zeta,\iota,\psi,\phi_0,\phi_{\alpha},\theta,\phi)A^i_{\alpha}(t,\omega_0),
\end{equation}
where,
\begin{align} \label{eq:aialpha}
a_{1\alpha} &= \left[q_{1\alpha}\left(1-\cos\phi_{\alpha}\right)-q_{2\alpha}\sin\phi_{\alpha}\right] \nonumber\\
a_{2\alpha} &= \left[q_{2\alpha}\left(1-\cos\phi_{\alpha}\right)+q_{1\alpha}\sin\phi_{\alpha}\right] \nonumber\\
q_{1\alpha} &= \left(F^+_{\alpha}a_1+F^{\times}_{\alpha}a_3\right)\nonumber\\
q_{2\alpha} &= \left(F^+_{\alpha}a_2+F^{\times}_{\alpha}a_4\right),\nonumber\\
\end{align} 
and,
\begin{align}
a_1 &= \zeta\left[\left(1+\cos^2\iota\right)\cos2\phi_0\cos2\psi + 2\cos\iota\sin2\phi_0\sin2\psi\right]  \nonumber\\
a_2 &= -\zeta\left[\left(1+\cos^2\iota\right)\sin2\phi_0\cos2\psi - 2\cos\iota\cos2\phi_0\sin2\psi\right]  \nonumber\\
a_3 &= \zeta\left[\left(1+\cos^2\iota\right)\cos2\phi_0\sin2\psi - 2\cos\iota\sin2\phi_0\cos2\psi\right]  \nonumber\\
a_4 &= -\zeta\left[\left(1+\cos^2\iota\right)\sin2\phi_0\sin2\psi + 2\cos\iota\cos2\phi_0\cos2\psi\right].
\end{align}

In the above equations, $\phi_{\alpha} = 2(\Phi_0-\Phi_{p\alpha,0})$ and $\zeta=\mathcal{M}^{5/3}/D_L$. The signal basis-functions are defined as,
\begin{equation}\label{eq:non-evolve-basis}
A^1_{\alpha} = \frac{1}{\omega_0^{1/3}}\sin(2\omega_0 t),\quad A^2_{\alpha} = \frac{1}{\omega_0^{1/3}}\cos(2\omega_0 t).
\end{equation}
We employ the log-likelihood ratio as a statistic for parameter-estimation and detection. This statistic is defined as the logarithm of the ratio of the likelihood of a signal being present to the signal being absent,
\begin{equation} \label{eq:log-like-ratio}
\ln\Lambda \equiv \ln p(\vec{r}|\vec{s}) - \ln p(\vec{r}|\vec{0}) = \sum_{\alpha=1}^{N_p} \left[(r_{\alpha}|s_{\alpha}) - \frac{1}{2}(s_{\alpha}|s_{\alpha})\right],
\end{equation}
where we have defined an inner product such that $(x|y) = x^T G(GCG)^{-1}G^T y$, where: (i) $C$ is a covariance matrix describing stochastic influences to the pulsar TOAs; (ii) $G$ is a timing-model marginalisation matrix \citep{van-haasteren-levin-2012}; and (iii) $r_{\alpha}$ is a vector of timing-residuals in the $\alpha^{\text{th}}$ pulsar. With well-constrained pulsar noise properties (fixed $C$) we can use $\Lambda$ within a Bayesian search to recover parameter posterior distributions. In this way we can also use $\Lambda$ to substitute for the full likelihood in the Bayesian evidence evaluation to recover the {\it Bayes factor}, allowing for a direct recovery of detection significance in a Bayesian way. Implicit in the derivation of Eq.\ (\ref{eq:log-like-ratio}) is the cancellation of the normalisation factor of the full likelihood (which is a function of $C$). This remains true when we form the evidence ratio, since we integrate over the deterministic signal parameter space and assume that stochastic noise properties are fixed. Explicitly,
\begin{align}
\mathcal{B} &= \frac{\mathcal{Z}_{\rm signal}}{\mathcal{Z}_{\rm null}}\nonumber\\
&= \frac{\int\exp{\left[ -(r-s(\vec\mu)|r-s(\vec\mu))/2\right]}\pi(\vec\mu)d^N\mu}{\int\exp{\left[ -(r|r)/2\right]}\pi(\vec\mu)d^N\mu} \\\nonumber
&= \frac{\int \exp{\left[ (r|s(\vec\mu)) - (1/2)(s(\vec\mu)|s(\vec\mu))\right]}\pi(\vec\mu)d^N\mu}{\int\pi(\vec\mu)d^N\mu} \\\nonumber
&= \int \Lambda(\vec\mu)\pi(\vec\mu)d^N\mu.
\end{align}

For the purposes of later analysis, we now write $\ln\Lambda$ explicitly in terms of the pulsar-phase parameters, $\phi_{\alpha}$. Defining
\begin{align}
N^i_{\alpha} &= (r_{\alpha}|A^i_{\alpha}),\nonumber\\
M^{ij}_{\alpha} &= (A^i_{\alpha}|A^j_{\alpha}),
\end{align}
and
\begin{equation}
\epsilon_i{}^j = \begin{pmatrix}0 & 1\\-1 & 0\end{pmatrix},
\end{equation}
such that
\begin{align}
\ln\Lambda& = \sum_{\alpha=1}^{N_p} \Big\{\left[q_{i\alpha}q_{j\alpha}M_{\alpha}^{ij} - q_{i\alpha}N_{\alpha}^i\right]\cos\phi_{\alpha} \nonumber\\
&+ \left[q_{k\alpha}q_{j\alpha}M_{\alpha}^{ij} - q_{k\alpha}N_{\alpha}^i\right]\epsilon_i{}^k\sin\phi_{\alpha} \nonumber\\
&- \frac{1}{2}q_{i\alpha}q_{j\alpha}M_{\alpha}^{ij}\cos^2\phi_{\alpha} - \frac{1}{2}q_{k\alpha}q_{l\alpha}M_{\alpha}^{ij}\epsilon_i{}^k\epsilon_j{}^l\sin^2\phi_{\alpha} \nonumber\\
&- q_{i\alpha}q_{k\alpha}M_{\alpha}^{ij}\epsilon_j{}^k\sin\phi_{\alpha}\cos\phi_{\alpha} + q_{i\alpha}N_{\alpha}^i - \frac{1}{2}q_{i\alpha}q_{j\alpha}M_{\alpha}^{ij}\Big\}.
\end{align}
where $N_p$ is the number of pulsars in our array. With negligible frequency evolution, the binary's parameters are $\left\{\zeta,\omega_0,\theta,\phi,\iota,\psi,\psi_0\right\}$, however we must also take into account an additional phase variable per pulsar, $\phi_{\alpha}$. Hence, in a parameter-estimation search or an evaluation of the Bayes factor, conventional techniques would require a search over $7+N_{p}$ dimensions. For large arrays or expensive likelihood evaluations this can be a costly exercise, necessitating multi-threading linear-algebra operations to accelerate the likelihood evaluations, or multi-core machines to perform efficient parallel-tempering for the evaluation of Bayes factors. One should also note that the popular and effective Bayesian inference tool \textsc{MultiNest} can struggle in these kinds of high-dimensional problems (even in constant efficiency mode) when we have complicated parameter spaces or lengthy likelihood evaluation times, as the set of live-points used in the nested sampling algorithm very slowly accumulates the last few units of log-evidence.\footnote{Certain alternative approaches to this have been proposed for \textsc{MultiNest}, such as the use of importance nested sampling in constant-efficiency mode \citep{importanceMultiNest2013}, or employing a trained neural network \citep{skynet} to accelerate the final stages of sampling.}

\subsection{Evolving template}

We can also write down an evolving-signal template which takes into account the orbital evolution of the SMBH binary during the Earth-pulsar light travel-time, but still assumes evolution during the actual PTA observation window is negligible. For this evolving-signal template, we define,
\begin{align}
A^1_{\alpha} = \frac{1}{\omega_0^{1/3}}\sin(2\omega_0 t),&\quad A^2_{\alpha} = \frac{1}{\omega_0^{1/3}}\cos(2\omega_0 t) \nonumber\\
B^1_{\alpha} = \frac{1}{\omega_{p,\alpha}^{1/3}}\sin(2\omega_{p,\alpha} t),&\quad B^2_{\alpha} = \frac{1}{\omega_{p,\alpha}^{1/3}}\cos(2\omega_{p,\alpha} t),
\end{align}
where $\omega_{p,\alpha} = \omega_0 - \dot{\omega_0}L_{\alpha}(1+\hat{\Omega}\cdot\hat{p}_{\alpha})$, $\dot\omega_0 = (96/5)\mathcal{M}^{5/3}\omega_0^{11/3}$, and $L_{\alpha}$ is the distance to the $\alpha^{\rm th}$ pulsar. 

In addition to $M^{ij}$ and $N^i$ for the non-evolving case, we define,
\begin{align}
O^{ij}_{\alpha} &= (B^i_{\alpha}|B^j_{\alpha}),\nonumber\\
P^i_{\alpha} &= (r_{\alpha}|B^i_{\alpha}),\nonumber\\
Q^{ij}_{\alpha} &= (A^i_{\alpha}|B^j_{\alpha}).
\end{align}

Now, expressing the log-likelihood ratio explicitly in terms of the pulsar-phase parameters, $\phi_{\alpha}$, gives the following,
\begin{align}
\ln\Lambda& = \sum_{\alpha=1}^{N_p} \Big\{\left[q_{i\alpha}q_{j\alpha}Q_{\alpha}^{ij} - q_{i\alpha}P_{\alpha}^i\right]\cos\phi_{\alpha} \nonumber\\
&+ \left[q_{k\alpha}q_{j\alpha}Q_{\alpha}^{ji} - q_{k\alpha}P_{\alpha}^i\right]\epsilon_i{}^k\sin\phi_{\alpha} \nonumber\\
&- \frac{1}{2}q_{i\alpha}q_{j\alpha}O_{\alpha}^{ij}\cos^2\phi_{\alpha} - \frac{1}{2}q_{k\alpha}q_{l\alpha}O_{\alpha}^{ij}\epsilon_i{}^k\epsilon_j{}^l\sin^2\phi_{\alpha} \nonumber\\
&- q_{i\alpha}q_{k\alpha}O_{\alpha}^{ij}\epsilon_j{}^k\sin\phi_{\alpha}\cos\phi_{\alpha} + q_{i\alpha}N_{\alpha}^i - \frac{1}{2}q_{i\alpha}q_{j\alpha}M_{\alpha}^{ij}\Big\}.
\end{align}

\subsection{Techniques for maximisation and marginalisation over $\phi_{\alpha}$}

By explicitly exposing $\phi_{\alpha}$ in our expressions for the likelihood-ratio, we have developed several alternative approaches designed to approximate maximisation or marginalisation of the likelihood-ratio over these pulsar-phase variables. 

Firstly, in the context of non-evolving templates, \citet{ellisoptimal2012} noted that one can avoid the formalism of the $\mathcal{F}_p$ statistic (which maximises the likelihood-ratio over $2N_{p}$ ``amplitude'' parameters [$a_{i\alpha}$ in Eq.\ (\ref{eq:non-evolve-template})] despite there being only $7+N_{p}$ independent parameters). 
Improving upon the $\mathcal{F}_p$ statistic is desirable, since as we expand the number of pulsars in our array the disparity between the dimensionality of the parameter-space assumed by the $\mathcal{F}_p$ statistic and the true {\it physical} parameter-space grows larger. Rather than maximising over these nuisance ``amplitude'' parameters, we can instead analytically maximise over the {\it physical} $\phi_{\alpha}$ parameters. This requires solving a quartic equation in $x=\cos\phi_{\alpha}$ which is guaranteed to have at least one unique solution, although whether that solution satisfies the requirement $-1 \leq x \leq 1$ must be ascertained on the fly. We can of course, avoid this completely by \textit{numerically} maximising over the pulsar-phase parameters. This is {\bf Technique $\#1$}, and constitutes a more appropriate maximisation than $\mathcal{F}_p$. Nevertheless, we are still left with the problem of searching over the remaining $7$-dimensional parameter space; this is a much more tractable problem and can be handled with many off-the-shelf MCMC or nested-sampling algorithms. In this case, we should not be surprised if a bias is observed in the posterior distributions of the final $7$ parameters, since we are after all \textit{maximising} over $N_p$ other parameters. 

The second option is to avoid maximising entirely, and instead marginalise the likelihood-ratio over the pulsar-phase parameters. Note that we can analytically marginalise over the amplitudes of the signal basis-functions in Eq.\ (\ref{eq:non-evolve-template}) with uniform-priors to get the Bayes factor for a common-frequency signal in pulsar TOAs. We do not discuss this further here, but provide the derivation and a brief analysis in Appendix \ref{sec:BpStatistic_sec}. The approach we follow here is to \textit{numerically} marginalise over the pulsar-phase parameters, such that we actually sample the marginalised likelihood-ratio in our MCMC or nested-sampling algorithms. In particular, if we can do this without increasing the likelihood evaluation time significantly, then the collapse of the dimensionality makes this problem tractable with \textsc{MultiNest}. There are many benefits to this; for example, \textsc{MultiNest} is an excellent tool for sampling multimodal distributions, it has inbuilt parallelisation, and in low-dimensionality provides an efficient means to evaluate the Bayesian evidence. Hence, the numerical marginalisation of the non-evolving template over pulsar-phase parameters is our {\bf Technique $\#2$}.

As a final point of interest for non-evolving templates, we note that if there are sufficiently many wave cycles during the observation time of the pulsars in our array, then it is possible to maximise over the pulsar-phase parameters analytically without the need to solve a quartic. More interestingly, it is also possible to analytically marginalise over the pulsar-phase parameters. The noise behaviour of real pulsars and the GW frequencies to which we are most sensitive will likely prohibit us from making the assumptions required to analytically maximise/marginalise in this fashion. However, we provide the derivation and a brief analysis in Appendix \ref{sec:AnalyticPhiMarg}, where we find that this analytic marginalisation may be able to place useful constraints on the values of $\zeta=\mathcal{M}^{5/3}/D_L$ and the orbital frequency of a SMBH binary, but sky-localisation and Bayesian evidence recovery is biased.

There are two ways to proceed with an evolving template, but both involve numerical marginalisation over the pulsar-phase parameters. In {\bf Technique $\#3$} we compute $\omega_{p,\alpha}$ by fixing $L_{\alpha}$ to its catalogued value, while in {\bf Technique $\#4$} we internally average over the prior distribution of $L_{\alpha}$ by drawing the distance used to compute $\omega_{p,\alpha}$ from a Gaussian centred on the catalogued value with standard-deviation given by the catalogued error-bars.

Even though the pulsar-phase has an explicit dependence on the pulsar-distance, including the distance in parameter estimation can produce practical difficulties, as a small change in the distance may have a relatively small effect on the pulsar-term frequency, $\omega_{p,\alpha}$, but can have a huge impact on the phase coherence \citep{corbin-cornish-2010,ellisbayesian2013}. Without sub-pc precisions on measured pulsar-distances the possibility of including the pulsar-term in a coherent analysis might seem beyond reach. However, \citet{ellisbayesian2013} overcomes this by sampling the distances on two scales; one is very small to maintain phase coherence, while the other is larger (on the order of kpc) to solve for the pulsar-term frequency. Regardless, highly-tuned jump proposals for any stochastic sampling approach seem necessary when trying to incorporate the pulsar-term in a coherent analysis. Our approximation side-steps this problem by marginalising over the pulsar-phase and drawing $L_{\alpha}$ from within its prior to calculate $\omega_{p,\alpha}$. We achieve significant accelerations with respect to the full search in two ways: (1) we perform an 8D search with a likelihood that executes $N_p\times$1D numerical integrations, as opposed to having to stochastically sample from an $(8+N_p)$D space; (2) this 8D search can be highly parallelised with \textsc{MultiNest} to minimise search times, as opposed to the lengthy burn-in times and prohibitive autocorrelation lengths associated with high-dimensional MCMC searches.

\section{Results} \label{sec:singlesourceresults}

\begin{table*}
\caption{\label{tab:9PsrInfo}Pulsar distances taken from \citet{verbiest-psr-distances} if available, or otherwise from the ATNF catalogue \citep{ATNF-cat}.}
\centering
\begin{tabular}{c c c c}
\hline
Pulsar & White-noise RMS [ns] & Time-span [yr] & Pulsar distances [kpc]\\
\hline
J0030+0451 & $792$ & $12.7$ & $0.28\pm0.1$ \\
J0437-4715 & $69$ & $14.8$ & $0.156\pm0.001$ \\
J1640+2224 & $410$ & $14.9$ & $1.19\pm0.238$ \\
J1713+0747 & $136$ & $18.3$ & $1.05\pm0.06$ \\
J1744-1134 & $366$ & $16.9$ & $0.42\pm0.02$ \\
J1857+0943 & $402$ & $14.9$ & $0.9\pm0.2$ \\
J1909-3744 & $100$ & $9.0$ & $1.26\pm0.03$ \\
J1939+2134 & $141$ & $16.3$ & $5.0\pm2.0$ \\
J2317+1439 & $412$ & $14.9$ & $1.89\pm0.38$ \\
\hline
\end{tabular}
\end{table*}

While a full analysis of these techniques in all conceivable situations is beyond the scope of this study, we rigorously test what we expect to be the most promising new technique. {\bf Technique $\#4$} (which from now we denote as the {\bf $\mathcal{M}_p$} statistic) is subjected to a program of systematic injection and recovery of simulated signals, using the \textsc{PALSimulation} code which is part of the \textsc{PAL} package \citep{PAL-site} being developed as a unifying suite of tools for pulsar timing analysis. The performance of {\bf Technique $\#3$} closely follows that of {\bf Technique $\#4$}, which is unsurprising since they involve similar methods. Furthermore, we expect no systematic bias from {\bf Technique $\#2$} other than that which is introduced by analysing an evolving signal within a non-evolving model.

The datasets we generated were of the following configurations;

\begin{itemize}

\item {\bf Type I:} $36$ pulsars, $5$ years of observations, $2$ week cadence, $100$ ns RMS white-noise per pulsar, $L_{\rm psr}=1\pm0.1$ kpc $\forall$ pulsars; equivalent to the assumptions of the first \textsc{Open} dataset in the IPTA MDC.

\item {\bf Type II:} $9$ pulsars, variable observation time-span, average $2$ week cadence, realistic white-noise, $L_{\rm psr}$ equal to catalogued values.

\item {\bf Type III:} $9$ pulsars, variable observation time-span, average $2$ week cadence, realistic white-noise, $L_{\rm psr}$ drawn from Gaussian distribution (mean$=$catalogued-value, standard-deviation$=$catalogued-error).

\end{itemize}

The observation time-spans, white-noise RMS values, and distances for the $9$ pulsars in Type II and Type III datasets are shown in Table \ref{tab:9PsrInfo}.

\subsection{Model selection}\label{sec:model-selection}

\begin{figure}
  \centering
   \subfloat[]{\incgraph{0}{0.5}{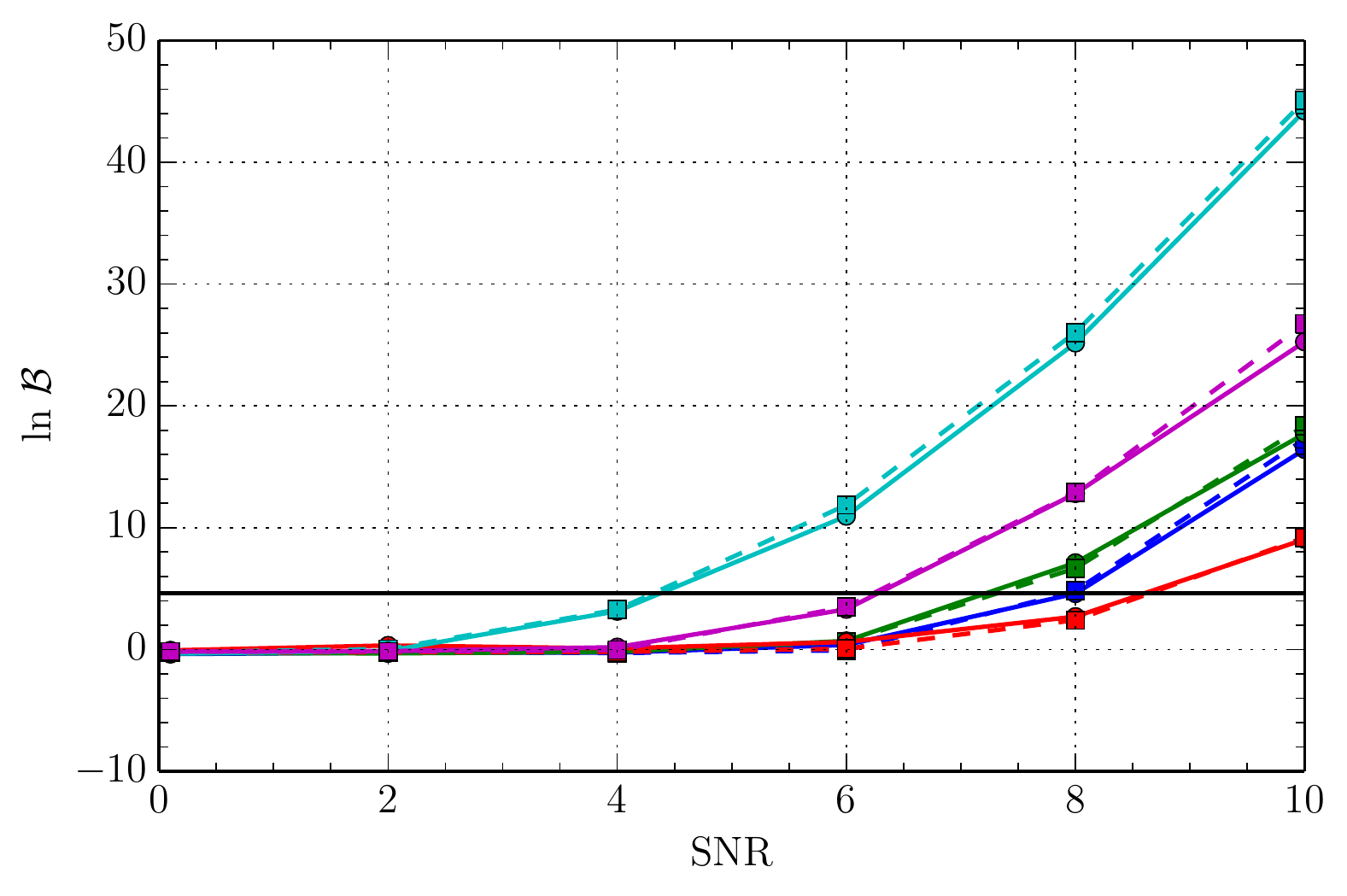}}\\
   \subfloat[]{\incgraph{0}{0.5}{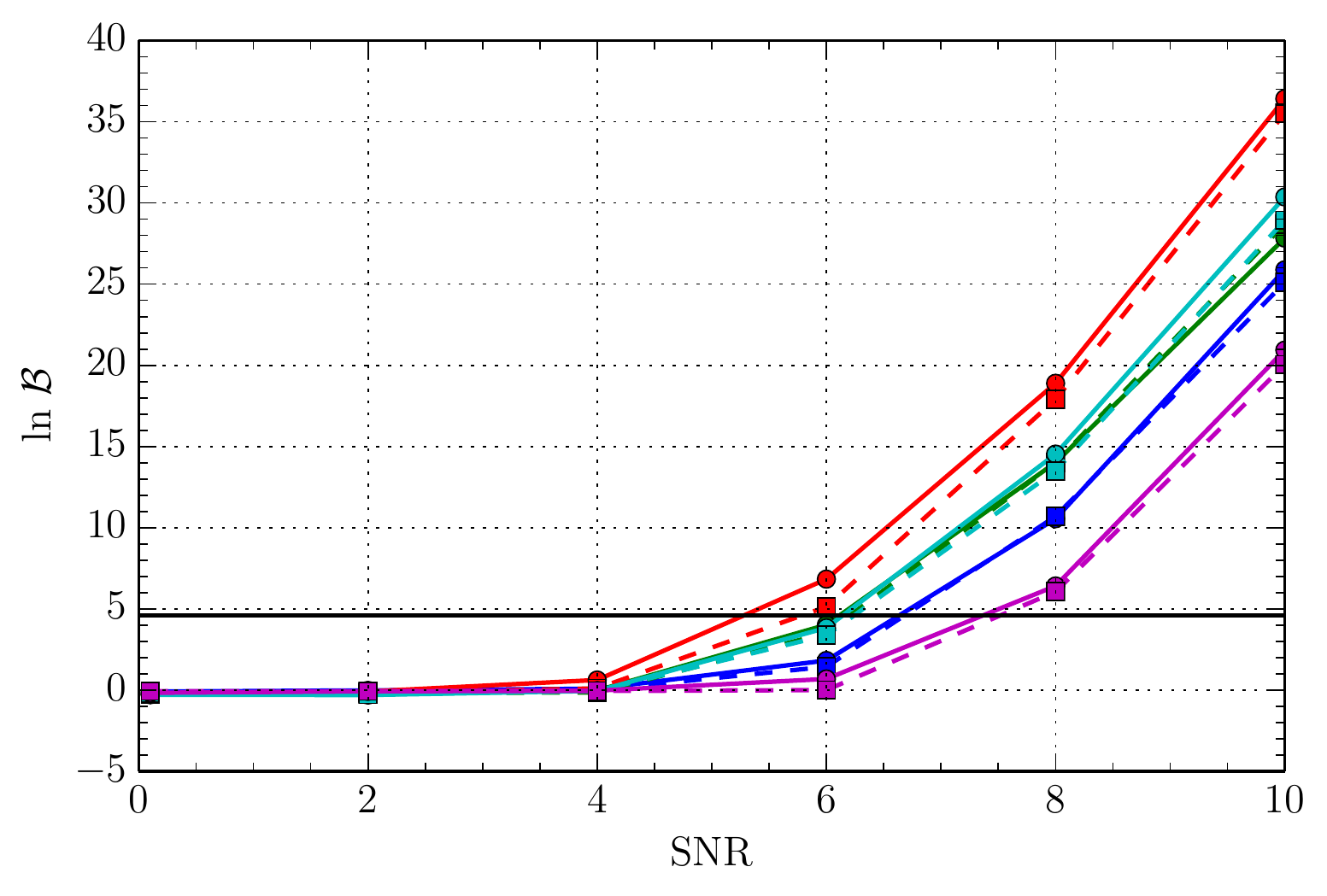}}
   \caption{\label{fig:typeIII_bayes}A comparison of the computed posterior odds-ratios ($\ln\mathcal{B}$) evaluated using thermodynamic integration of the full signal model (solid lines), and the technique of numerically marginalising over the pulsar-phase parameters while sampling from the pulsar-distance prior ({\bf $\mathcal{M}_p$} statistic; dashed lines). Different SNR signals are injected into a variety of realisations of Type I and Type III datasets. The agreement found between the two methods is excellent.} 
 \end{figure}

\begin{figure}
  \centering
\incgraph{0}{0.5}{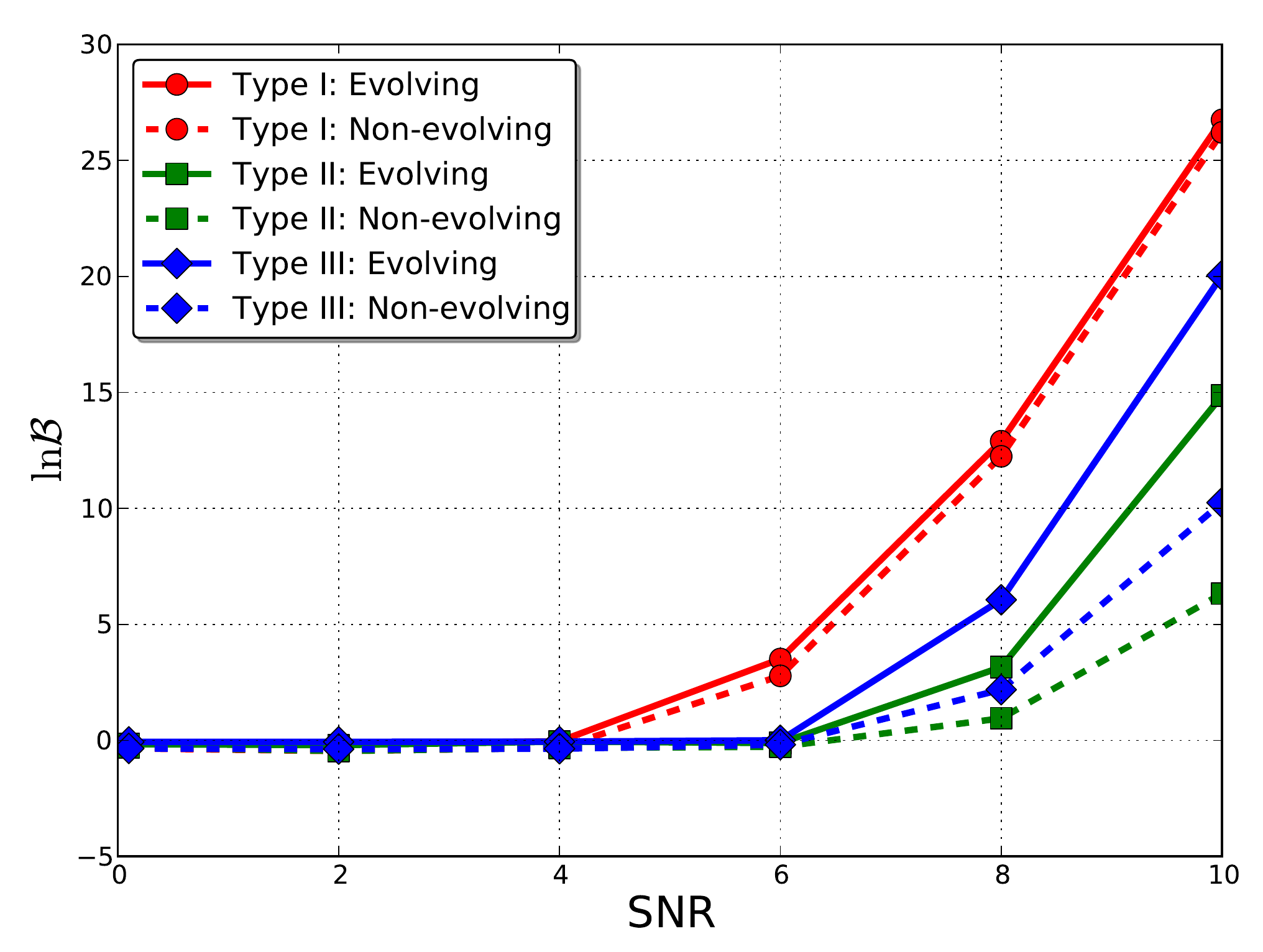}
\caption{\label{fig:EvNonEv_BayesCompare}For a given realisation of noise, we repeat the analysis of Type I/II/III datasets with a non-evolving template. We see that the mismatch between the assumption of a non-evolving signal and the reality of an evolving-binary injection leads to Bayes factors which can be significantly below the optimal evolving-model values.} 
 \end{figure}

We evaluate the accuracy of the Bayes factors returned by these pulsar-phase marginalisation techniques by injecting signals into various noise realisations at various SNRs. The SNR in these cases is defined as ${\rm SNR}^2 = \sum_{\alpha}\left(s\left(\vec{\mu}_{\rm inj}\right)|s\left(\vec{\mu}_{\rm inj}\right)\right)_{\alpha}$. We compare the recovered Bayes factors with those obtained by employing parallel-tempering and thermodynamic integration with the full signal template (and searching for the pulsar-distances). Parallel tempering is a method of launching many MCMC chains of varying ``temperature'' designed to aggressively search parameter space, and avoid trapping of chains in local likelihood maxima. Each chain has a different target distribution, $p(\vec\mu|D,\beta)\propto p(\vec\mu)p(D|\vec\mu)^{\beta}$, where $\beta$ is the inverse temperature and varies between $0$ and $1$. Higher temperatures effectively flatten out the likelihood surface, and explore regions far from maximum. A multi-temperature Hastings ratio is employed to ensure mixing of the chains and rapid localisation of the global maximum. After exploration the different chains can be processed via thermodynamic integration to give an estimate of the Bayesian evidence \citep[e.g.,][]{littenberg-cornish-2010}. The evidence for a chain with inverse temperature $\beta$ is simply,
\begin{equation}
\mathcal{Z}_{\beta} = \int\;d\vec\mu\; p(\vec\mu)p(D|\vec\mu)^{\beta},
\end{equation}
such that,
\begin{align}
\ln\mathcal{Z} &= \int_0^1d\beta\;\frac{\partial\ln\mathcal{Z}_{\beta}}{\partial\beta} \nonumber\\
&= \int_0^1d\beta\int\;d\vec\mu\; \frac{p(\vec\mu)p(D|\vec\mu)^{\beta}}{\mathcal{Z}_{\beta}}\ln p(D|\vec\mu)\nonumber\\
&= \int_0^1d\beta\;\langle\ln p(D|\vec\mu)\rangle_{\beta},
\end{align}
where $\langle\cdot\rangle_{\beta}$ denotes an expectation value with respect to the target posterior of inverse temperature $\beta$. For details on the parallel tempering and thermodynamic integration techniques employed here, see \citet[]{ellisbayesian2013, nanogravCW_2014} and references therein.


The signal we inject matches that explored in \citet{ellisbayesian2013}, which is at the sky-location of the Fornax cluster. Recent work has shown that there may be potential single GW source ``hot spots'' in the Virgo, Fornax and Coma clusters \citep{fornax-hotspot}. Regardless, we are only interested in sensible parameters to form an injected signal. These parameters are $\left\{\mathcal{M},D_L,f_0,\phi,\cos\theta,\cos\iota,\psi,\phi_0\right\} = \left\{7\times10^8M_{\odot},-,10^{-8}{\rm Hz},0.95,-0.56,0,1.26,2.65\right\}$, where the luminosity distance $D_L$ is scaled to suit the desired SNR. 

Another important aspect is our choice of prior on $\mathcal{M}$, $D_L$ and $f_0$. We employ log-uniform priors on these variables, but also apply a cut on the characteristic-strain induced by the binary, where we define $h_0 = 4\sqrt{2/5}\omega_0^{2/3}\zeta$ and require $h_0 \leq h_{0,c}\left(f_{\rm gw} / 10^{-8}{\rm Hz}\right)^{2/3}$, where $h_{0,c}=10^{-13}$. We use Monte Carlo integration to compute the prior re-normalisation, which only leads to a change in log odds-ratio of $\lesssim 0.1$. However, this cut had practical value in limiting the high-strain parameter space which was inhibiting our thermodynamic integration from converging to the true evidence value with a reasonable number of temperature chains. Also, this is a cheap way to impose a correlated prior on chirp mass, luminosity distance and GW frequency \citep{nanogravCW_2014}.  

The comparison between an evaluation of the posterior odds ratio performed by the full thermodynamic integration (solid lines) and the {\bf $\mathcal{M}_p$} statistic (dashed lines) for Type I and Type III datasets of various injected SNR is shown in Fig.\ \ref{fig:typeIII_bayes}, where we see excellent agreement for a variety of different noise realisations. For realistic Type III datasets, we in fact see that the {\bf $\mathcal{M}_p$} statistic gives a mildly conservative estimate of the full Bayes factor. We find that the speed of the numerical-marginalisation techniques depend on the SNR of the injection, where for low to moderate SNR ($\sim 0 - 2$) the evidence and parameter-estimation stages of \textsc{MultiNest} completed within {\it only a few minutes} of wall-time on $48$ computational cores. The highest SNR injections (SNR $=10$) required longer, but still finished within $\sim 45$ minutes of wall-time on $48$ cores. The reason for this trend is that the likelihood at low SNR is broad and featureless in the pulsar-term phase parameters, allowing the numerical integration routines to converge rapidly to a solution. In comparison, thermodynamic integration took more than a day for a single dataset analysis with similar computational resources. 



Analysing these datasets using the numerical phase marginalisation with a non-evolving template ({\bf Technique $\#2$}), we find that the mismatch between the model and the evolving-signal injections leads to Bayes factors which can be significantly below the optimal evolving-model values. This is illustrated for a single noise realisation in Figure \ref{fig:EvNonEv_BayesCompare}. We will revisit this in the next section.

\subsection{Parameter estimation}\label{sec:param-estimate}

\begin{figure}
  \centering
  \incgraph{0}{0.5}{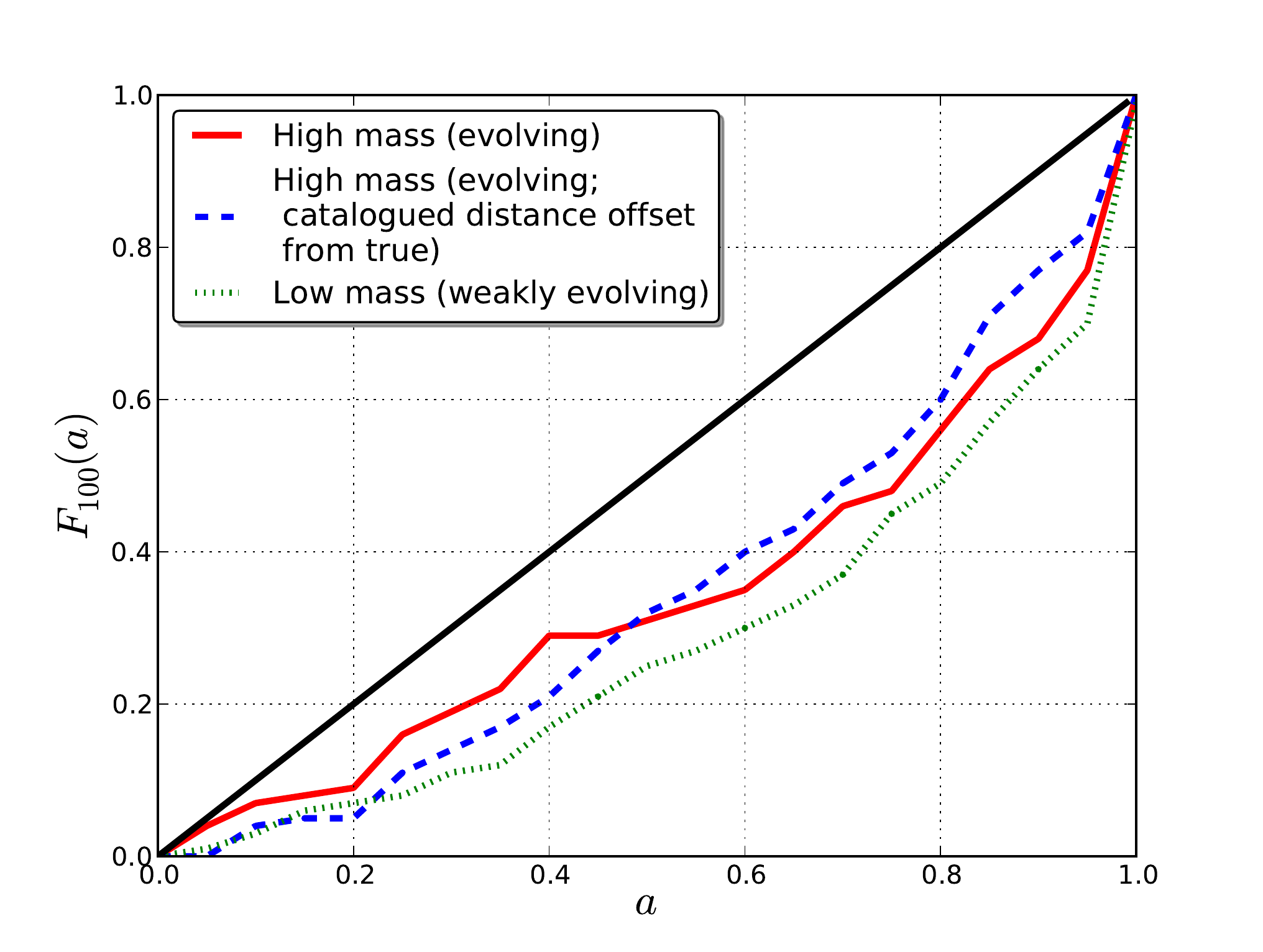}
   \caption{\label{fig:T4pp}The fraction of injections which are ``closer'' in the chi-squared sense (see text) to the set of points lying inside credible-interval, $a$, is plotted against the credible-interval. The line of zero-bias is shown as a thick, black-line, while the results of an analysis of $100$ realisations of evolving/non-evolving Type II datasets using numerical-marginalisation (the {\bf $\mathcal{M}_p$} statistic) are shown as solid-red and dashed-blue. The dashed-green line shows the result for when we offset our catalogue of distances from their true values by an amount consistent with their error-bars. While some bias is present, this plot does not indicate how that manifests in the physical parameter-space.} 
 \end{figure}

To ascertain whether numerical marginalisation techniques introduce any systematic bias in parameter recovery, we inject SNR $=8$ signals into various white-noise realisations. The injected binary orbital frequency is chosen to be $10^{-8}$ Hz such that the GW frequency lies close to the peak sensitivity of an array of pulsars observed over a period of $\gtrsim 5$ years (see \citet{moore-2014} for a full discussion of Bayesian and frequentist continuous-wave sensitivity curves, and \citet{nanogravCW_2014} for the latest NANOGrav continuous-wave sensitivity curves.). 
We choose injected chirp masses of $7\times 10^8M_{\odot}$ and $1.8\times 10^8M_{\odot}$ in order to model a strongly evolving (over the Earth-pulsar light-travel time), and weakly evolving binary respectively, where the lower mass injection will have an $\dot\omega$ which is $\sim 10\%$ of the higher mass.

These evolving and weakly-evolving binaries are injected into $100$ different noise-realisations of Type II datasets. This type of dataset is used because we want the characteristics of the PTA to remain fixed, such that the injected binary's luminosity distance, $D_L$, (which is scaled to accommodate the desired SNR) is constant over each realisation. The remaining binary parameters are injected with the following values into each dataset; $\left\{\phi=1,\cos\theta=0.48,\cos\iota=0.88,\psi=0.5,\right.$ $\left.\phi_0=2.89\right\}$. 

We present results for the case of the {\bf $\mathcal{M}_p$} statistic, which should be applicable regardless of whether the binary is evolving or not. We again note that no bias would be expected within {\bf Technique $\#2$}, which numerically marginalises over the pulsar-phase variables in the non-evolving formalism. The only bias expected here derives from the inherent limitations of applying an inappropriate non-evolving model to a possibly evolving signal. 



\begin{figure*}
  \centering
\incgraph{0}{1.0}{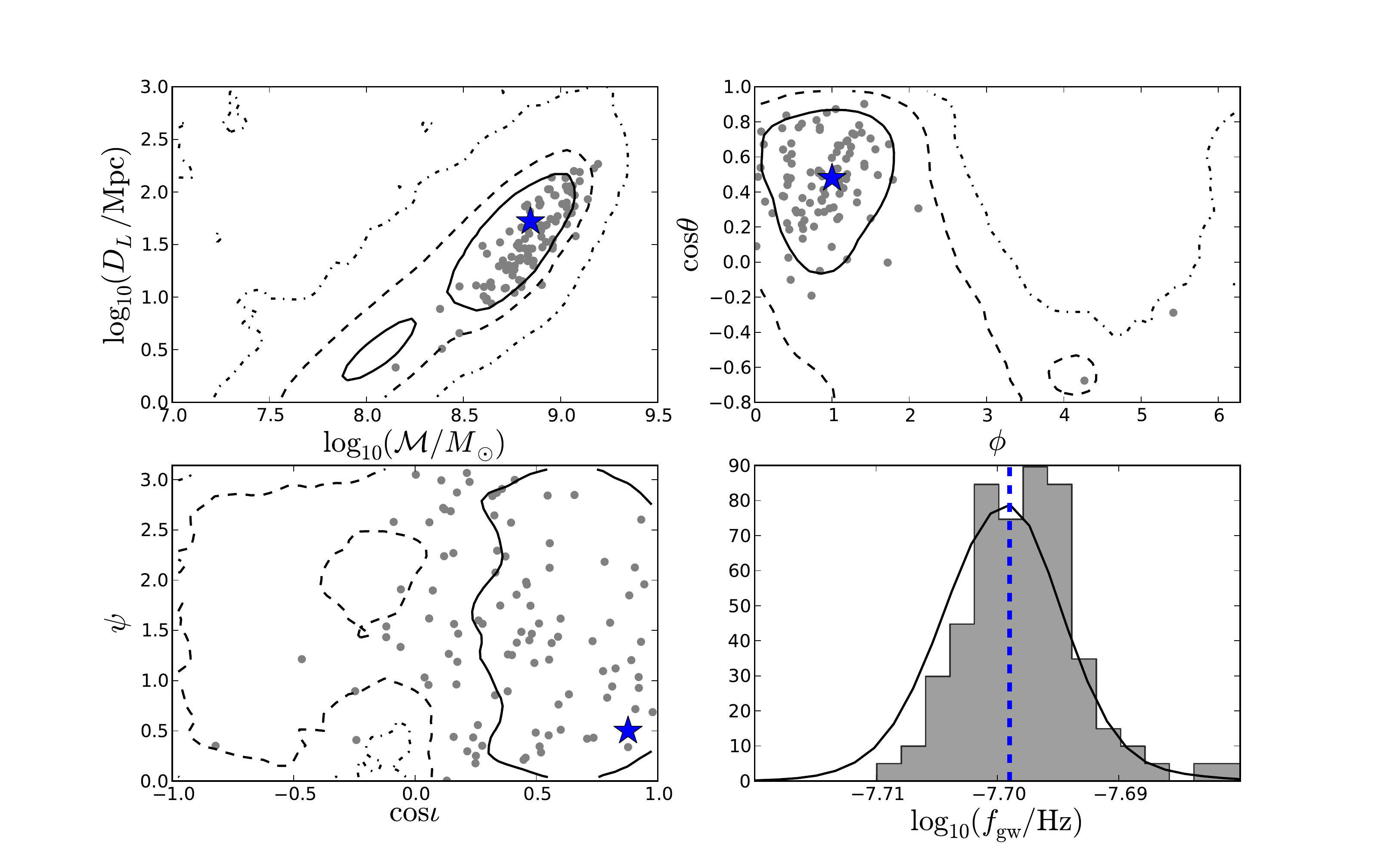}
\caption{\label{fig:MaxPost2}
We show the distribution of {\it maximum-a-posteriori} values (filled grey circles) from an analysis of $100$ realisations of an evolving signal injected into a Type II dataset, and analysed with the {\bf $\mathcal{M}_p$} statistic. As a further step towards real dataset analysis, we offset our catalogue of pulsar distances from their true values by an amount consistent with error bars. As can be seen, in the parameters of interest $(\mathcal{M},D_L,f_{\rm gw},\phi,\cos\theta,\cos\iota)$ this technique recovers the injected values (blue stars and blue dashed lines) quite comfortably. Additionally, we overplot the $68\%$, $95\%$ and $99\%$ contours of the posterior probability distributions averaged over all noise realisations. On average the distribution of the maximum-a-posterior values follows the average posterior, except perhaps in the case of $\cos\iota$. All injected values lie within the $68\%$ credible interval.} 
 \end{figure*}

\begin{figure*}
  \centering
\incgraph{0}{1.0}{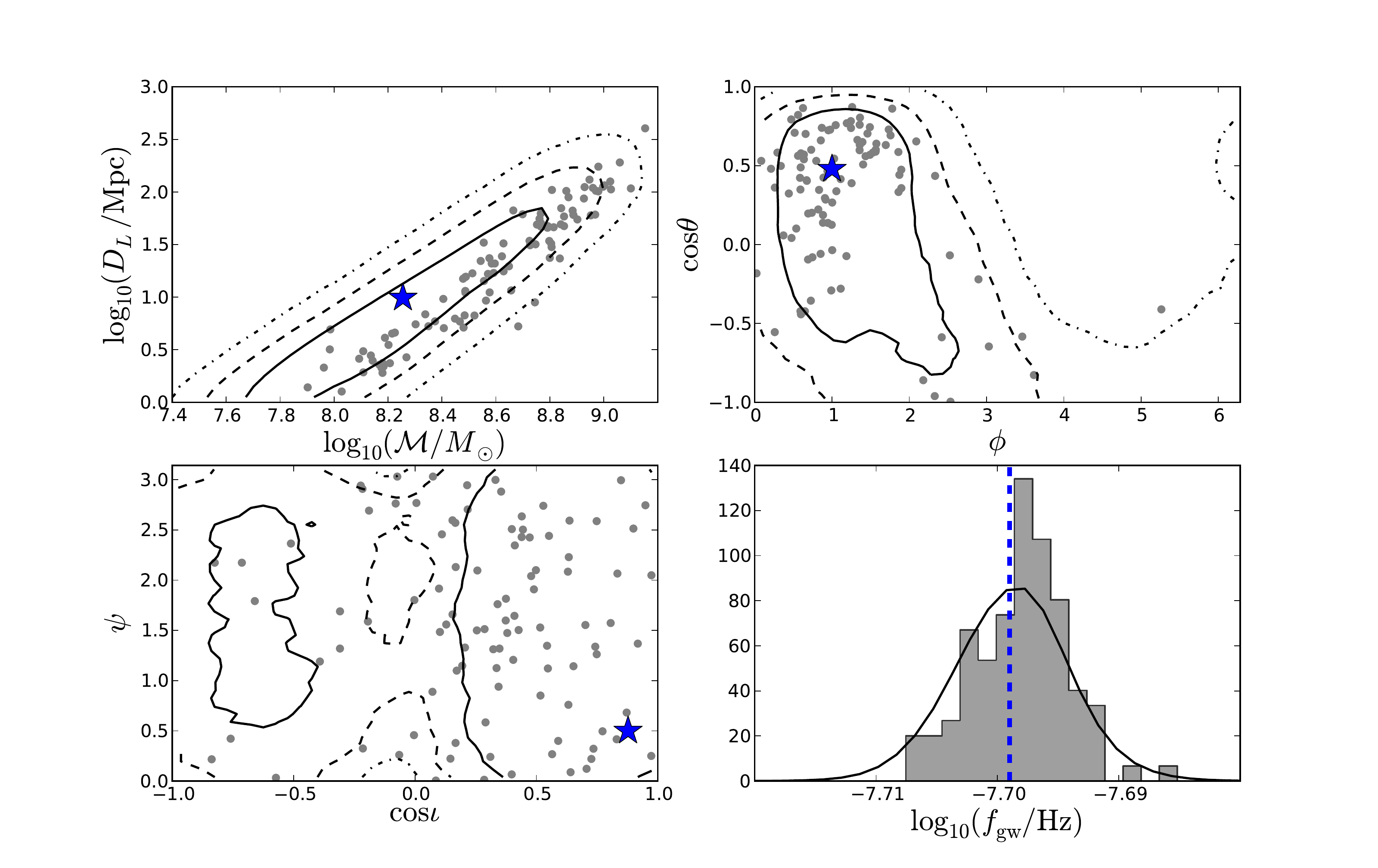}
\caption{\label{fig:MaxPost3}We show the distribution of {\it maximum-a-posteriori} values (filled grey circles) from an analysis of $100$ realisations of a weakly evolving signal injected into a Type II dataset, and analysed with the {\bf $\mathcal{M}_p$} statistic. The injected values of $(\mathcal{M},D_L)$ appear to be offset from the distribution of {\it maximum-a-posteriori} values, but are fully consistent with the overplotted average posterior probability distributions (see Fig.\ \ref{fig:MaxPost2} for additional details). We note that all injected values lie within the $68\%$ credible interval.}
 \end{figure*}

\begin{figure}
  \centering
  \incgraph{0}{0.5}{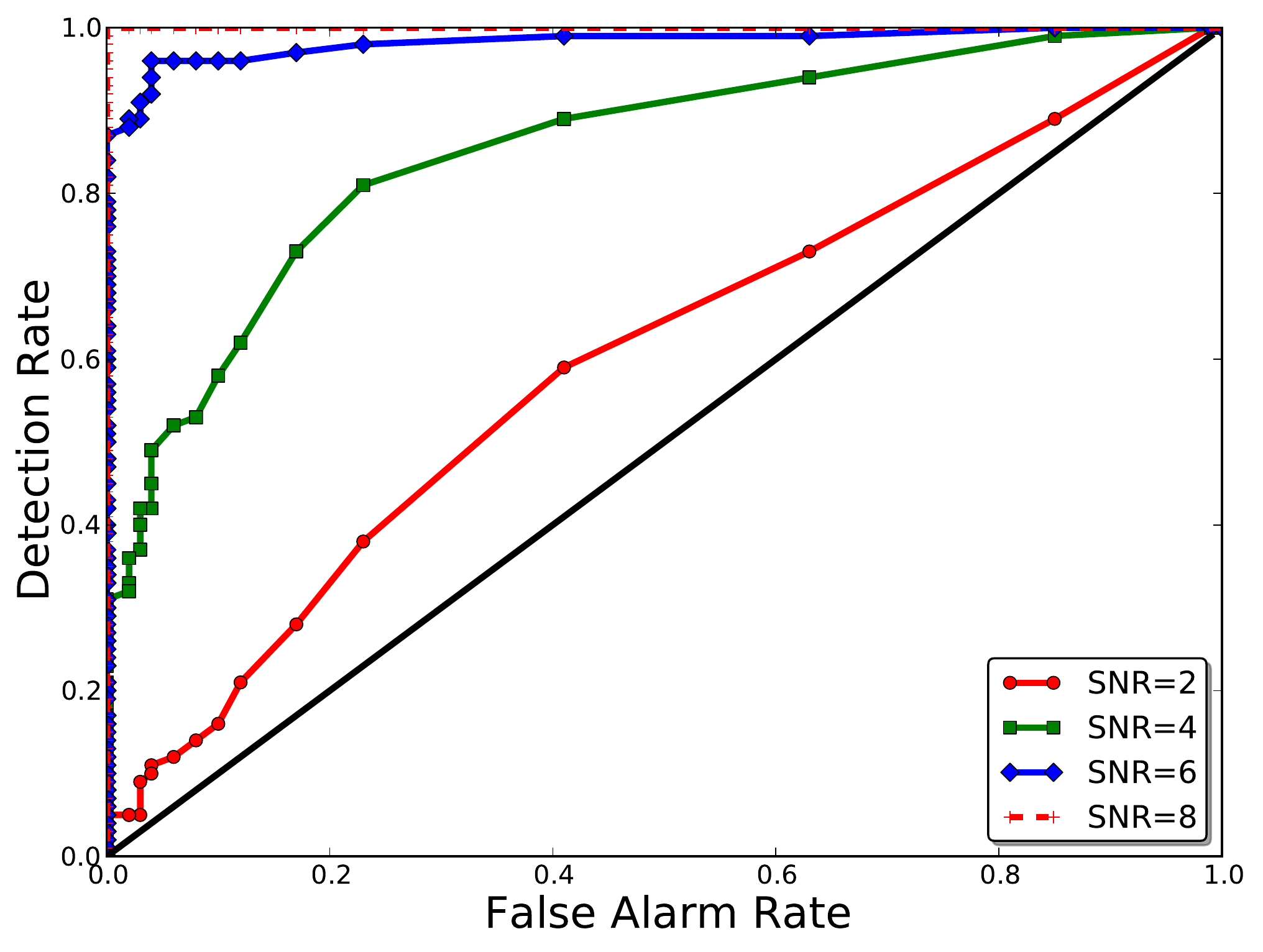}
   \caption{\label{fig:ROC}We inject an evolving signal into $100$ realisations of Type II datasets at various SNRs (including SNR=0), recovering the posterior-odds ratio via the {\bf $\mathcal{M}_p$} statistic in each case. Setting the threshold of detection at varying values of the posterior odds ratio, we compute the fraction of realisations which are classified as false-positive and true-positive detections. We see that for this binary, and using this technique, the posterior odds ratio is an almost perfect classifier at SNR=6. With these numerical marginalisation techniques, the run-time is fast enough to permit detailed analysis of detection requirements within a Bayesian context.} 
 \end{figure}

Our method of testing for systematic bias in the use of the {\bf $\mathcal{M}_p$} statistic is an extension of a method used in \citet{ellis-first-order} to validate the accuracy of a first-order likelihood approximation in a stochastic background search. As discussed there, the benchmark of internal consistency is when, in $x\%$ of realisations, the set of injected parameters lies within the inner $x\%$ of the marginalised posterior distribution. The inner high-probability region is defined as,
\begin{align}
\int_W p(\vec\theta )d^N\theta &= a, \\\nonumber
W = \{\theta^1,\theta^2,\ldots,\theta^N\in\mathbb{R} &: p(\vec\theta )>\mathcal{L}_a\},
\end{align}
where $\mathcal{L}_a>0$ is some value unique to each $a$ corresponding to a curve of equal probability in the $N$ dimensional parameter space.

To find all points satisfying $p(\vec\theta )>\mathcal{L}_a$  we rank the recovered posterior samples in order of decreasing posterior weight, then integrate over all samples until we reach the desired credible interval. For each realisation, we can then define two sets of points; the set of points inside the high-probability region (HPR) $\mathcal{S}_a$, and the complementary set $\mathcal{S}_{\bar a}$.

We now extend the dimensionality of the definitions of the $\chi^2$ variables in \citet{ellis-first-order} to give a measure of the distance of the posterior samples in each set from the true injected parameters,
\begin{align}
&\chi_a(\vec\theta_i)^2 = \left(\frac{\log_{10}(\mathcal{M}_i)-\log_{10}(\mathcal{M}_{\rm true})}{\log_{10}(\mathcal{M}_{\rm true})}\right)^2 \nonumber\\
&+ \left(\frac{\log_{10}(D_{L,i})-\log_{10}(D_{L,\rm true})}{\log_{10}(D_{L,\rm true})}\right)^2 + \left(\frac{\phi_i-\phi_{\rm true}}{\phi_{\rm true}}\right)^2 \nonumber\\ 
&+ \left(\frac{\cos\theta_i-\cos\theta_{\rm true}}{\cos\theta_{\rm true}}\right)^2 + \left(\frac{\cos\iota_i-\cos\iota_{\rm true}}{\cos\iota_{\rm true}}\right)^2 \nonumber\\
&+ \left(\frac{\psi_i-\psi_{\rm true}}{\psi_{\rm true}}\right)^2 + \left(\frac{\phi_{0,i}-\phi_{0,\rm true}}{\phi_{0,\rm true}}\right)^2, \\\nonumber
\end{align}
where $\vec\theta_i$ are elements of $\mathcal{S}_a$. We also define a corresponding expression for $\chi_{\bar a}(\vec\theta_j)^2$ in terms of the elements, $\vec\theta_j$, of the complementary set, $\mathcal{S}_{\bar a}$.

Finally, we define the empirical distribution function (EDF) as,
\begin{equation}
F_k(a) = \frac{1}{k}\sum_{n=1}^k\Theta\left({\rm min} \chi_{\bar a}^2 - {\rm min} \chi_a^2\right),
\end{equation}
where $k$ is the number of noise realisations, and $\Theta(x)$ is the Heaviside step-function. This summation gives the fraction of all noise-realisation in which the injected values are ``closer'' (in the $\chi^2$ sense) to one of the elements of the HPR than to any element of the complementary set. 

The results of such an analysis are shown in Fig.\ \ref{fig:T4pp} for the evolving and weakly-evolving binary injections. The line of internal consistency is shown as a thick, black diagonal line. We see that this technique does indeed present bias, with a worst-case sag of $\sim 0.25$. However, the EDF does not give an insight into how this bias manifests itself in the parameter space. 

In Fig.\ \ref{fig:MaxPost2} we show the distribution of {\it maximum-a-posteriori} values over all $100$ noise-realisations, with the injected signal parameters also indicated. It is clear that while the {\bf $\mathcal{M}_p$} statistic may fail the formal EDF test, in practical terms it quite comfortably recovers the true parameters of the injected signal. This holds even when the catalogue of pulsar distances is offset from the true values by an amount consistent with their error-bars. Additionally, we show how the injected parameters and maximum-a-posteriori values are distributed with respect to the $68\%$, $95\%$ and $99\%$ contours of the realisation-averaged posterior. On average the distribution of the maximum-a-posterior values follows the average posterior, except perhaps in the case of $\cos\iota$, which may be the source of the bias seen in the formal EDF test. Regardless, all injected values lie within the $68\%$ credible interval. 
The {\bf $\mathcal{M}_p$} statistic also recovers the true injected parameter values when the GW source is weakly evolving. Figure \ref{fig:MaxPost3} shows a similar analysis to Fig.\ \ref{fig:MaxPost2} for a weakly evolving injection, where, despite some offset of the injected values of $(\mathcal{M},D_L)$ from the distribution of maximum-a-posteriori values, all injected values lie within the $68\%$ credible interval of the overplotted realisation-averaged posterior probability distributions.

A further test we carry out is to assess the performance of the $\mathcal{M}_p$-computed Bayesian posterior odds-ratio as a detection classifier. We do so by producing a receiver operator characteristic (ROC) plot, illustrating the fraction of true positive detections versus false positive detections as we vary the detection threshold. We inject various SNR signals into $100$ different noise-realisations, recovering the evidence in each case. The injected binary parameters are the same as the evolving case above. We see from Fig.\ \ref{fig:ROC} that the posterior odds-ratio becomes a virtually perfect detection classifier at an SNR of $6$. Although we cannot draw truly general conclusions from this, the aim of this exercise is to show that these numerical marginalisation techniques are accurate enough to allow detailed statistical tests within a Bayesian context with much lower computational expenditure than existing techniques. The question of what is required for an unambiguous claim of GW detection using Bayesian statistics has been hitherto out of reach due to high computational expenditure, but can be rigorously assessed by employing these techniques.

\begin{figure}[h]
  \centering
  \incgraph{0}{0.5}{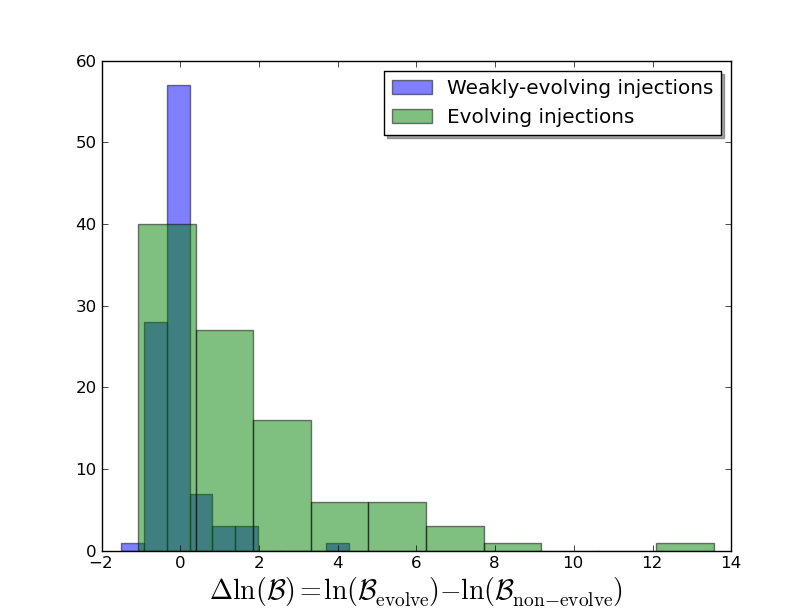}
   \caption{\label{fig:EvNonEv_BayesDiff}Evolving and weakly-evolving signals are injected into $100$ different noise-realisations with an SNR of $8$. We analyse all datasets using both the non-evolving and evolving templates (with numerical phase marginalisation), recovering the evidence in each case. We find that evidence recovered using the numerical phase marginalisation conforms to expected behaviour. On average, when the injected signal is weakly-evolving there is no difference in the evidence for an evolving or non-evolving template. However, when the signal is evolving the distribution of evidence will on average favour the evolving template.}
 \end{figure}

Finally, we assess the importance of using an evolving versus non-evolving template when establishing detection criteria. For evolving and weakly-evolving sources, we inject SNR=8 signals into $100$ different noise-realisations. We analyse each dataset using both the numerical phase marginalisation in the evolving-model ($\mathcal{M}_p$ statistic) and the non-evolving model, recovering the evidences in each case. The results are shown in Fig.\ \ref{fig:EvNonEv_BayesDiff}, where we see that the evolving template is more general, capturing the behaviour of the gravitational-waveform even when the signal is non-evolving, and giving a Bayes factor which is comparable to the value returned by the non-evolving analysis. However, as seen in the previous section, the non-evolving template recovers a Bayes factor which can be significantly lower than the evolving-model template whenever the signal is truly evolving. 
This shows that the evidence values returned by these numerical phase marginalisation techniques conform to expected behaviour, and allow us to infer whether the GW signal is evolving based on the evolving versus non-evolving posterior odds ratio.

\section{Conclusion} \label{sec:singlesourceconclusions}

Near-future GW searches which exploit the high-precision timing of millisecond pulsars may open a new observational window onto the early-inspiral phase of SMBH binaries. These systems are expected to be ubiquitous in the current picture of hierarchical structure formation, where massive galaxies grow via accretion from cosmic web filaments and galactic mergers \citep{whiterees1978,kauffmann2000}. Supermassive BHs are thought to reside within the nuclei of most galaxies \citep[e.g.,][]{ferrarese2005}, evolving symbiotically with the host \citep[e.g.,][]{ferrarese2000,magorrian1998,marconi2003}, such that galactic mergers, followed by the inspiral of BHs via dynamical friction into the post-merger remnant, leave a large population of SMBH binary systems.

While the dominant nanohertz GW signal accessible to PTAs will likely be a stochastic background formed from the incoherent superposition of signals from the inspiral of these systems, massive nearby binaries may be visible as single resolvable sources. Detecting these systems, and determining their properties, will offer a complementary probe to eLISA/NGO of the massive BH-population, in addition to a cross-check of system parameters from possible electromagnetic counterparts \citep[see][and references therein]{burkespolaor2013}. These counterparts may in fact aid detection, as we no longer need to perform completely blind searches and can collapse the parameter space of our search algorithms.

In this paper we have presented several new approaches to single-source searches in PTAs. The need to include the pulsar-term in analyses for accurate sky-localisation leads to practical difficulties, as distances to pulsars are poorly constrained, requiring us to introduce an extra search-parameter per pulsar. In evolving-template searches we must also take into account the inspiral of the binary over Earth-pulsar light travel-times, which (when we coherently include the pulsar-term) effectively extends the baseline of our observations by thousands of years, allowing our searches to reconstruct the orbital-evolution of the system and disentangle its chirp mass from the luminosity distance.

By numerically marginalising ``on-the-fly'' over the phase of the GW as it passes each pulsar, and sampling the distance to each pulsar from prior electromagnetic constraints, we can collapse the dimensionality of our searches. Our likelihood is fast enough, and our search space small enough, to bring the powerful Bayesian inference package \textsc{MultiNest} to bear on the problem. We achieve significant accelerations with respect to the full search in two ways: (1) we perform an 8D search with a likelihood that executes $N_p\times$1D numerical integrations, as opposed to having to stochastically sample from an $(8+N_p)$D space; (2) this 8D search can be highly parallelised with \textsc{MultiNest} to minimise search times, as opposed to the lengthy burn-in times and prohibitive autocorrelation lengths associated with high-dimensional MCMC searches. For low to moderate SNRs we can perform parameter-estimation and recover the Bayesian evidence within a few minutes, whereas a full search utilising thermodynamic integration can take as long as a day with similar computational resources. We find excellent agreement of our Bayes factors with those returned by full searches, and, although the parameter estimation shows some small level of systematic bias in formal EDF tests, in practical terms we quite comfortably recover injected parameters. Analytic marginalisation of the likelihood over the pulsar-term phases may be able to place useful constraints on the values of $\zeta=\mathcal{M}^{5/3}/D_L$ and the orbital frequency of a SMBH binary, although sky-localisation and Bayesian evidence recovery is biased. 

We will apply these techniques to upcoming continuous GW searches with EPTA and IPTA datasets. Our techniques are fast enough to allow systematic injection and recovery of many signals, permitting an exploration of the criteria required to make an unambiguous Bayesian detection claim.

\appendix
\section{Analytic marginalisation and maximisation over $\phi_\alpha$ in non-evolving template} \label{sec:AnalyticPhiMarg}
\begin{figure*}
  \centering
   \subfloat[]{\incgraph{0}{0.5}{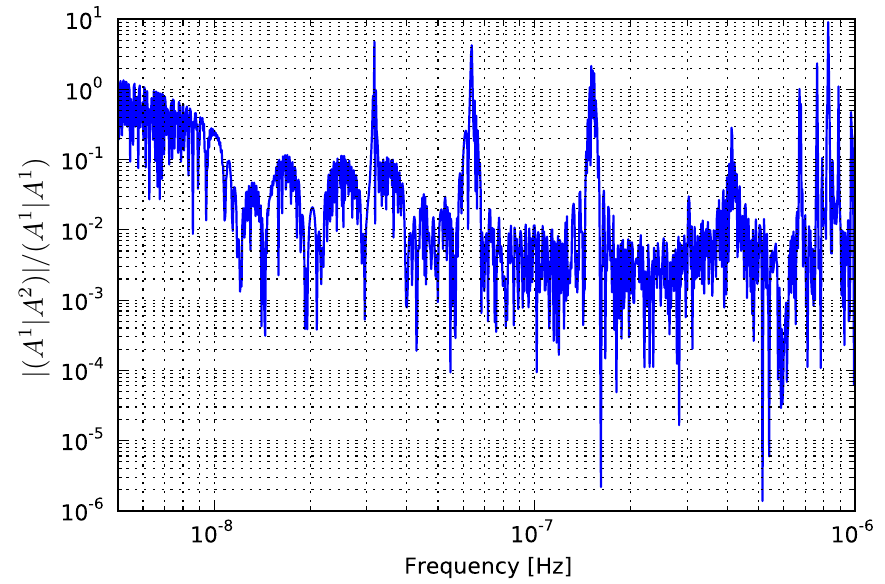}} 
   \subfloat[]{\incgraph{0}{0.5}{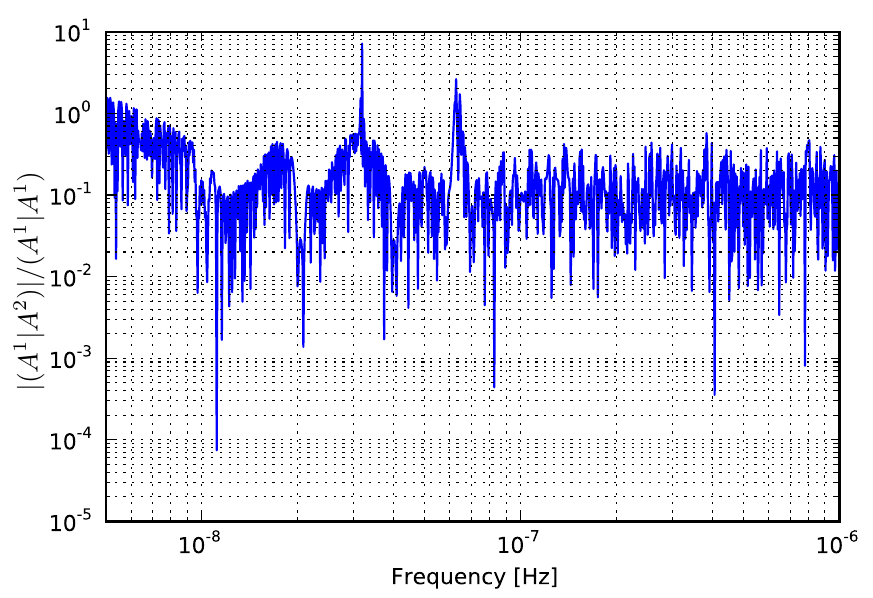}}
   \caption{\label{fig:overlap_ratios}The ratio of the basis-function overlaps in the cross-terms and the diagonal terms, $(A^1|A^2)/(A^1|A^1)$, is shown for (a) an IPTA MDC Open1 pulsar; 100 ns RMS white-noise, 2 week cadence; (b) a real NANOGrav dataset for J0613-0200 \citep{demorest-2012}, where the noise is also fairly white.} 
 \end{figure*}

In the following we refer to the non-evolving template of Sec.\ \ref{sec:NonEvolveTemplate}. Assuming we have sufficiently many wave cycles during the observation time-span, we can use the following assumptions for the signal basis-function overlaps in Eq.\ (\ref{eq:non-evolve-basis}): $(A^1|A^2)=(A^2|A^1)\simeq 0$, and $(A^1|A^1)\simeq(A^2|A^2)\simeq \mathcal{N}(\omega_0)$. In practice, the ratio of the cross-terms of the basis-function overlaps to the diagonal terms may not be small enough to permit these approximations to be used. For example, Fig.\ \ref{fig:overlap_ratios} shows the ratio $(A^1|A^2)/(A^1|A^1)$ for one of the pulsars in the IPTA MDC Open1 dataset, and for a real NANOGrav J0613-0200 dataset \citep{demorest-2012}. The ratio diminishes at higher frequencies, and for the mock dataset gets to $\lesssim 10^{-2}$ at the highest detectable frequencies. However, for a real pulsar dataset the ratio stays around $10^{-1}$ even at the highest frequencies. Furthermore, the GW frequencies to which we are most sensitive are $\sim$ a few $\times 10^{-8}$, diminishing as we move to the higher frequencies required for these approximations to hold.

Nevertheless, these analytic expressions may have some value as rapid first-pass tools, and we provide the derivations below.
 
\subsection{Marginalising}

Given the overlap approximations and the non-evolving template defined in Eq.\ (\ref{eq:non-evolve-template}-\ref{eq:non-evolve-basis}) we have,
\begin{align}
(s_{\alpha}|s_{\alpha}) &\simeq \left[a_{1{\alpha}}a_{1{\alpha}} + a_{2{\alpha}}a_{2{\alpha}}\right]\mathcal{N}(\omega_0), \nonumber\\
&\simeq 2\mathcal{N}(\omega_0)\left(q_{1\alpha}^2+q_{2\alpha}^2\right)\left(1-\cos\phi_{\alpha}\right),
\end{align}
such that,
\begin{align} \label{eq:collapsedlnlambda}
\ln\Lambda &= \sum_{\alpha=1}^{N_p} \left[(r_{\alpha}|s_{\alpha}) - \frac{1}{2}(s_{\alpha}|s_{\alpha})\right]  \nonumber\\
&\simeq \sum_{\alpha=1}^{N_p} \left\{\left[q_{1\alpha}(r_{\alpha}|A^1_{\alpha})+q_{2\alpha}(r_{\alpha}|A^2_{\alpha}) - \left(q_{1\alpha}^2+q_{2\alpha}^2\right)\mathcal{N}(\omega_0)\right]\right. \nonumber\\
&- \left[q_{1\alpha}(r_{\alpha}|A^1_{\alpha})+q_{2\alpha}(r_{\alpha}|A^2_{\alpha}) - \left(q_{1\alpha}^2+q_{2\alpha}^2\right)\mathcal{N}(\omega_0)\right]\cos\phi_{\alpha}  \nonumber\\
&\left. - \left[q_{2\alpha}(r_{\alpha}|A^1_{\alpha})-q_{1\alpha}(r_{\alpha}|A^2_{\alpha})\right]\sin\phi_{\alpha}\right\}\nonumber\\
&\simeq \sum_{\alpha=1}^{N_p}\left[ -X_{\alpha} + X_{\alpha}\cos\phi_{\alpha} + Y_{\alpha}\sin\phi_{\alpha}\right].
\end{align}

Hence, marginalising the likelihood-ratio over each pulsar-phase parameter, assuming flat-priors, gives,
\begin{align}
\int\Lambda\;d^{N_p}\phi &\propto \left(\frac{1}{2\pi}\right)^{N_p}\prod_{\alpha=1}^{N_p}\int_0^{2\pi}\exp[(r_{\alpha}|s_{\alpha})-\frac{1}{2}(s_{\alpha}|s_{\alpha})]d\phi_{\alpha} \nonumber\\
&\propto \left(\frac{1}{2\pi}\right)^{N_p}\exp\left(-\sum_{\alpha=1}^{N_p}X_{\alpha}\right) \nonumber\\
&\times\prod_{\alpha=1}^{N_p}\int_0^{2\pi}\exp(X_{\alpha}\cos\phi_{\alpha}+Y_{\alpha}\sin\phi_{\alpha})\;d\phi_{\alpha} \nonumber\\
&\propto \prod_{\alpha=1}^{N_p}\exp(-X_{\alpha})I_0\left(\sqrt{X_{\alpha}^2+Y_{\alpha}^2}\right),
\end{align}
where $I_0$ is a modified Bessel function of the first kind. Note that this technique of analytic marginalisation of nuisance phase parameters has previously been used in different contexts \citep{whalen1971detection,F-and-G}, but has never been applied to PTA data-analysis. Finally, we have the PML (Phase Marginalised Likelihood) statistic,
\begin{equation}
\ln\tilde\Lambda \propto \sum_{\alpha=1}^{N_p}\left\{-X_{\alpha} + \ln\left[I_0\left(\sqrt{X_{\alpha}^2+Y_{\alpha}^2}\right)\right]\right\}.
\end{equation}

If we have a high SNR signal, such that the argument of the modified Bessel function is large, then directly computing $I_0(x)$ can be very difficult. However, we can use a large argument expansion of the modified Bessel function to aid this calculation,
\begin{align}
 \ln\left[I_0(x)\right] &\sim x-\frac{1}{2}\ln\left(2\pi x\right) + \ln\left(1+\frac{1}{8x}+\frac{9}{128x^2}\right. \nonumber\\
&\left. +\frac{225}{3072x^3}+\frac{11025}{98304x^4}\ldots\right).
\end{align}

\begin{figure*}
  \centering
  \subfloat[]{\incgraph{0}{0.75}{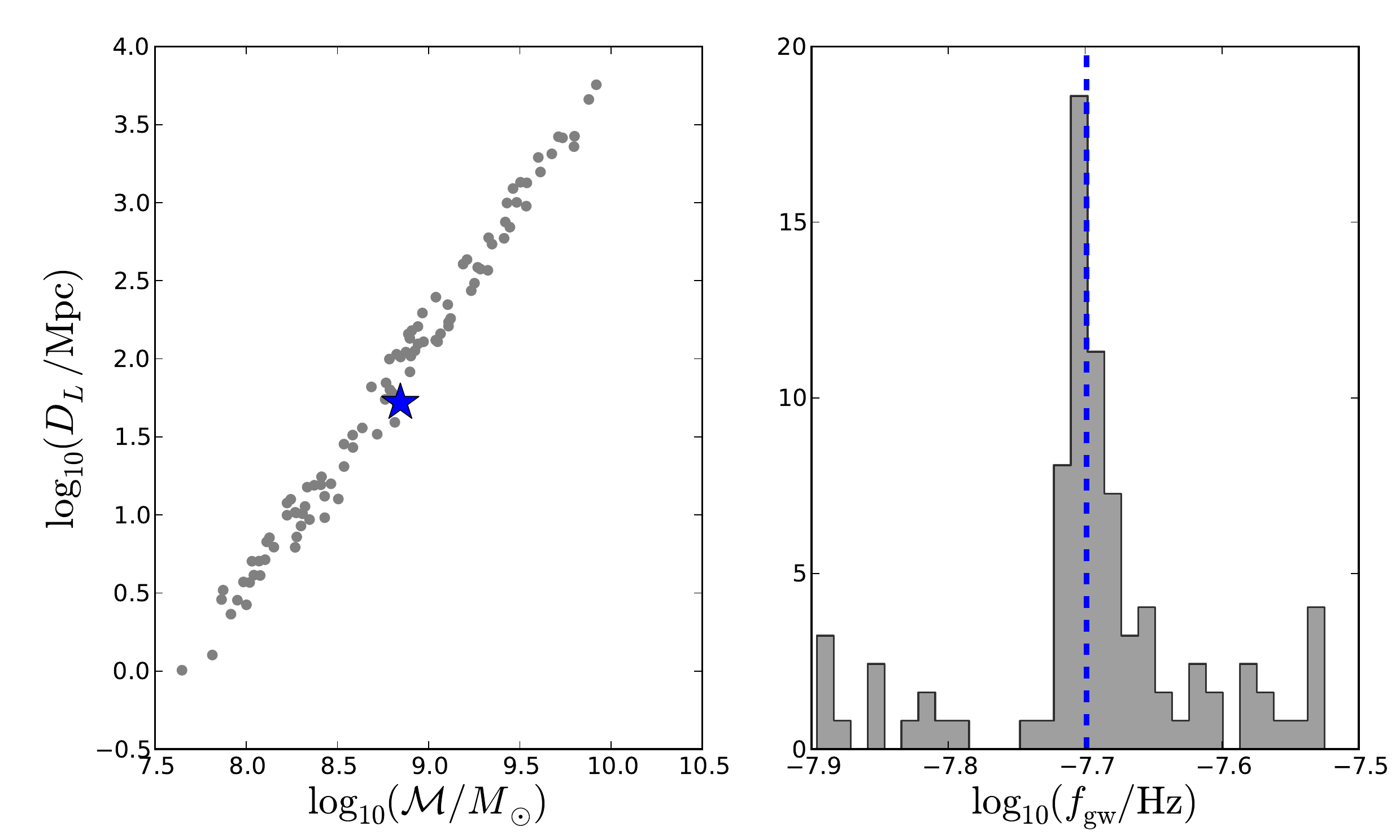}}\\
  \subfloat[]{\incgraph{0}{0.75}{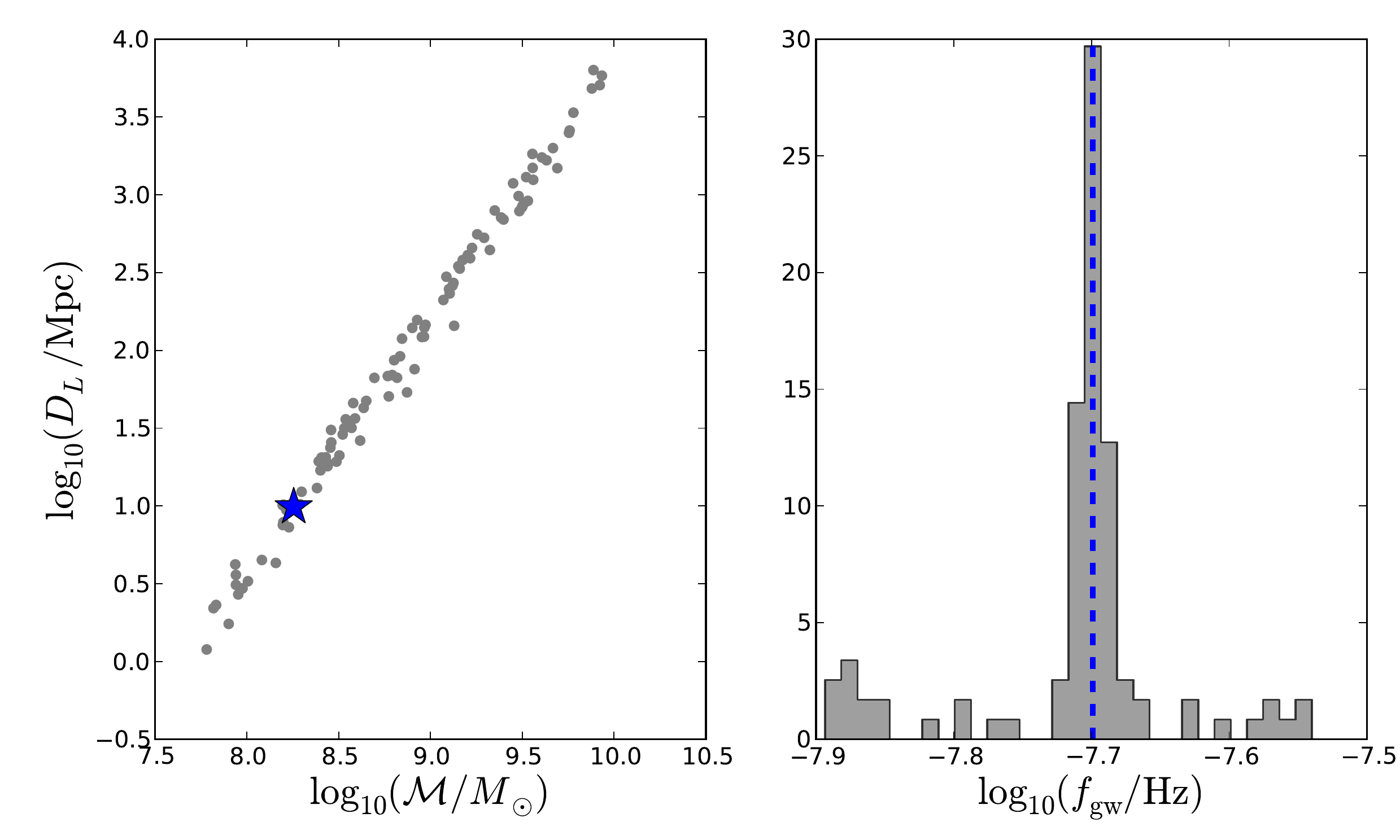}}
   \caption{\label{fig:MAP_NonEvAnalaytic}We show the distribution of {\it maximum-a-posteriori} values (filled grey circles) from analyses of $100$ realisations of (a) evolving and (b) weakly-evolving signals injected into Type II datasets (see Sec.\ \ref{sec:param-estimate} for details). These datasets were analysed with the Phase Marginalised Likelihood (PML) statistic, which involves an analytic marginalisation over pulsar-term phase parameters. In both cases the injected values (blue stars and blue dashed lines) of $\mathcal{M}$, $D_L,$ and $f_{\rm gw}$ are consistent with the distribution of maximum-a-posteriori values, however other parameters showed significant bias.} 
 \end{figure*}

We applied this statistic to the SNR=8 evolving and weakly-evolving datasets discussed in Sec.\ \ref{sec:param-estimate}. The analysis proceeded very quickly with minimal computational resources, since we are only searching over $8$ parameters without any expensive stages in the likelihood evaluation. In Fig.\ \ref{fig:MAP_NonEvAnalaytic} we show the distribution of maximum-a-posteriori values from the analysis of $100$ noise realisations. The injected values of $\mathcal{M}$, $D_L,$ and $f_{\rm gw}$ are consistent with the distribution of maximum-a-posteriori values, however other parameters showed significant bias. The recovered Bayes factors were also highly biased. Hence the PML statistic may be useful in placing constraints on the binary's $\zeta=\mathcal{M}^{5/3}/D_L$ and orbital frequency, although sky-localisation and Bayesian evidence recovery is unreliable.

\subsection{Maximising}

Going back to the original $\ln\Lambda$ in Eq.\ (\ref{eq:collapsedlnlambda}), it is possible to maximise the likelihood-ratio over the pulsar-phase parameters. As indicated in \citet{ellisoptimal2012}, the solution to the maximum-likelihood value of $\phi
_{\alpha}$ requires evaluating a quartic. However, if we use the overlap approximations from the previous section then the solution is more simple. Maximising gives
\begin{equation}
\frac{\partial\ln\Lambda}{\partial\phi_{\beta}} \simeq -X_{\beta}\sin\phi_{\beta} + Y_{\beta}\cos\phi_{\beta}=0,
\end{equation}
where
\begin{equation}
\tan\phi_{\beta}=\frac{Y_{\beta}}{X_{\beta}},
\end{equation}
so that we can define the log-likelihood ratio maximised over all $\phi_{\alpha}$, which we call the $\mathcal{T}_p$-statistic,
\begin{equation}
\mathcal{T}_p = \sum_{\alpha=1}^{N_p}\left[-X_{\alpha} + \sqrt{X_{\alpha}^2+Y_{\alpha}^2}\right].
\end{equation}

We may be able to go further, and to maximise over other parameters, but we do not consider this here. Regardless, we have a rather compact form for the log-likelihood ratio maximised over all the pulsar-phase parameters. The remaining $7$-D single-source parameter space can easily be explored using MCMC.

Note that if we use the large argument expansion of the modified Bessel function to approximate the PML we get,
\begin{equation}
\ln\tilde\Lambda \propto \sum_{\alpha=1}^{N_p}\left\{-X_{\alpha} + \sqrt{X_{\alpha}^2+Y_{\alpha}^2}-\frac{1}{2}\ln\left(2\pi\sqrt{X_{\alpha}^2+Y_{\alpha}^2}\right)\right\}.
\end{equation}

For sufficiently large arguments, $\sqrt{X_{\alpha}^2+Y_{\alpha}^2}$ increases faster than $\ln\left(2\pi\sqrt{X_{\alpha}^2+Y_{\alpha}^2}\right)$. Hence, in the infinite SNR limit the PML statistic is proportional to the maximum-likelihood estimator $\mathcal{T}_p$ statistic,
\begin{equation}
\ln\tilde\Lambda \propto \sum_{\alpha=1}^{N_p}\left\{-X_{\alpha} + \sqrt{X_{\alpha}^2+Y_{\alpha}^2})\right\} \propto\mathcal{T}_p.
\end{equation}

\section{$\mathcal{B}_p$ statistic (analytic marginalisation over amplitude parameters in non-evolving template)} \label{sec:BpStatistic_sec}

\begin{figure*}
  \centering
   \incgraph{0}{0.8}{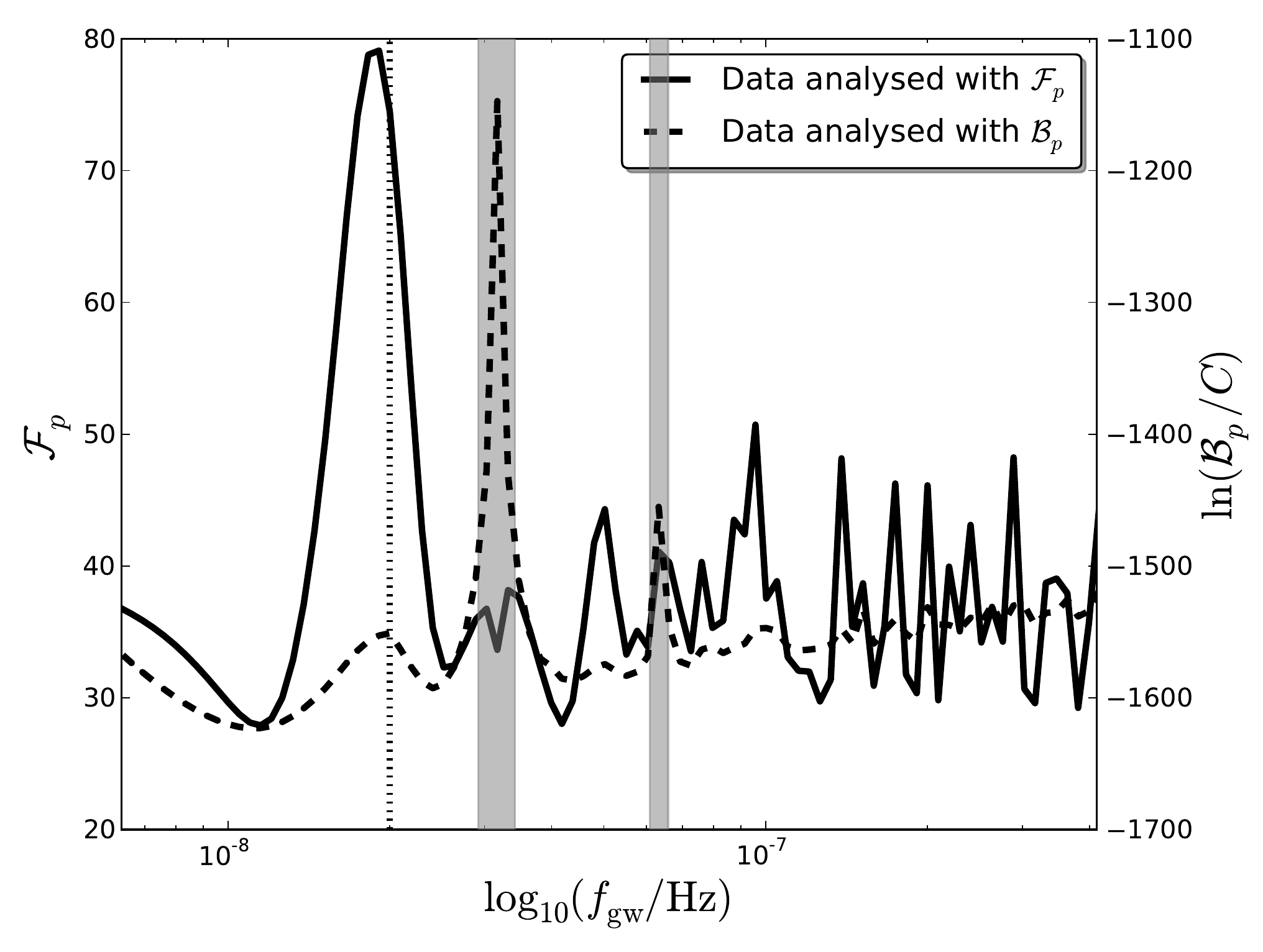}
   \caption{\label{fig:BpStat}A Type I dataset with an SNR=10 injection (with injected parameters equal to those in Sec.\ \ref{sec:model-selection}) was analysed with the $\mathcal{F}_p$ statistic and the $\mathcal{B}_p$ statistic. The injected GW frequency is $2\times 10^{-8}$ Hz and is shown as a dotted line. The $\mathcal{F}_p$ statistic performs very well and finds the true signal frequency. The $\mathcal{B}_p$ statistic also shows a small peak at this frequency, however the extra determinant factor in Eq.\ (\ref{eq:bp}) leads the shape of the frequency trend to closely resemble the noise curve. The grey regions around $\sim 6.34\times 10^{-8}$ Hz and $\sim 3.17\times 10^{-8}$ Hz correspond to a loss of sensitivity of the PTA due to conversion of topocentric TOAs to barycentric TOAs and fitting for parallax, respectively.} 
 \end{figure*}

Rather than analytically maximising over the amplitude parameters, $a_{i\alpha}$ [see Eq.\ (\ref{eq:aialpha})], to produce the $\mathcal{F}_p$ statistic, if we assume uniform priors on these parameters then it is trivial to analytically marginalise and calculate the Bayes factor. We re-write the likelihood as the following and complete the square in the amplitude parameters, such that
\begin{align}
\ln\Lambda &= \sum_{\alpha=1}^{N_p} (r_{\alpha}|s_{\alpha}) - \frac{1}{2}(s_{\alpha}|s_{\alpha}) \nonumber\\
=& \sum_{\alpha=1}^{N_p} a_{i\alpha}(r_{\alpha}|A^i_{\alpha}) - \frac{1}{2}a_{i\alpha}a_{j\alpha}(A^i_{\alpha}|A^j_{\alpha}) \nonumber\\
=& \sum_{\alpha=1}^{N_p} a_{i\alpha}N^i_{\alpha} - \frac{1}{2}a_{i\alpha}a_{j\alpha}M^{ij}_{\alpha}, \nonumber\\
=& -\frac{1}{2}\sum_{\alpha=1}^{N_p}\left[\left(a_{\alpha}-M_{\alpha}^{-1}N_{\alpha}\right)^T M_{\alpha}\left(a_{\alpha}-M_{\alpha}^{-1}N_{\alpha}\right)\right. \nonumber\\
&\left. - N_{\alpha}^T \left(M_{\alpha}^{-1}\right)^T N_{\alpha}\right].
\end{align}
Now we integrate over the amplitude parameters with uniform priors, and permit the maximum strain to be large enough such that the likelihood is unaffected by the prior boundary. We can therefore set the limits of integration to be between $[-\infty,+\infty]$, such that 
\begin{align}\label{eq:bp}
\mathcal{B}_p &= C\exp\left(\sum_{\alpha=1}^{N_p}\frac{N_{\alpha}^T \left(M_{\alpha}^{-1}\right)^T N_{\alpha}}{2}\right)\prod_{\alpha=1}^{N_p}\left[\text{det}\left(2\pi M_{\alpha}^{-1}\right)\right]^{1/2} \nonumber\\
&= C\left(2\pi\right)^{N_p}\exp\left(\mathcal{F}_p\right) \prod_{\alpha=1}^{N_p} \left(\text{det}\;M_{\alpha}\right)^{-1/2},
\end{align}
where $C$ denotes the prior volume. 

In Fig.\ \ref{fig:BpStat} we show the results of an application of the $\mathcal{B}_p$ statistic to a Type I dataset with an injected GW frequency equal to $2\times 10^{-8}$ Hz. The $\mathcal{F}_p$ statistic performs very well and unambiguously locates the correct signal frequency. While the $\mathcal{B}_p$ statistic also shows a small peak at this frequency, the extra determinant factor in Eq.\ (\ref{eq:bp}) causes the trend in frequency to show significant features of the noise curve. Hence, in this isolated case, $\mathcal{B}_p$ significantly underperforms $\mathcal{F}_p$. 

The form of the $\mathcal{B}_p$ statistic has been previously arrived at in the context of LIGO data analysis \citep{prix_bp}, where uniform priors for $a_{i\alpha}$ was shown to be very unphysical, and more physically-motivated priors were suggested. This was further explored in \citet{whelan2014}, where a new set of coordinates was found which are linear combinations of $a_{i\alpha}$, but which have a closer relationship to the physical parameter space. This improved the accuracy of the approximate analytic Bayes factor calculation with respect to the full numerical result. We do not explore this coordinate transformation here, but will consider this promising route in future work.






\acknowledgements
ST acknowledges the support of the STFC and the RAS. This research was in part supported by an appointment to the NASA Postdoctoral Program at the Jet Propulsion Laboratory, administered by Oak Ridge Associated Universities through a contract with NASA. JE is an Einstein fellow and acknowledges support by NASA through Einstein Fellowship grant PF4-150120. JE was partially funded through an NSF CAREER award number 0955929 and through the Wisconsin Space Grant Consortium. JG is supported by the Royal Society. Part of this work was performed using the Darwin Supercomputer of the University of Cambridge High Performance Computing Service (http://www.hpc.cam.ac.uk/), provided by Dell Inc.\ using Strategic Research Infrastructure Funding from the Higher Education Funding Council for England. Part of the computational work was performed on the Nemo cluster at UWM supported by NSF grant number 0923409.

\bibliography{MargPhase_Refs}

\end{document}